\documentclass[aip, amsmath,amssymb,reprint]{revtex4-1}
\usepackage{graphicx}
\usepackage{dcolumn}
\usepackage{bm}
\usepackage[utf8]{inputenc}
\usepackage[T1]{fontenc}
\usepackage{mathptmx}
\usepackage[dvipsnames]{xcolor}
\usepackage[toc,page]{appendix} 
\usepackage{tabularx}
\usepackage{url}
\usepackage{amsfonts}
\usepackage[mathscr]{eucal}

\let\oldAA\AA
\renewcommand{\AA}{\text{\normalfont\oldAA}}

\begin{document}

\preprint{AIP/123-QED}

\title{Protein--Polymer Mixtures in the Colloid Limit: Aggregation, Sedimentation and Crystallization}

\author{Rui Cheng}
\affiliation{HH Wills Physics Laboratory, Tyndall Avenue, Bristol, BS8 1TL, UK}
\affiliation{Bristol Centre for Functional Nanomaterials, University of Bristol, Tyndall Avenue, Bristol, BS8 1TL, UK}

\author{Jingwen Li}
\affiliation{HH Wills Physics Laboratory, Tyndall Avenue, Bristol, BS8 1TL, UK}
\affiliation{Bristol Centre for Functional Nanomaterials, University of Bristol, Tyndall Avenue, Bristol, BS8 1TL, UK}
\affiliation{School of Chemistry, University of Bristol, Cantock's Close, Bristol, BS8 1TS, UK.}

\author{Ioatzin R\'ios de Anda}
\affiliation{HH Wills Physics Laboratory, Tyndall Avenue, Bristol, BS8 1TL, UK}
\affiliation{School of Mathematics, University Walk, Bristol, BS8 1TW, UK.}

\author{Thomas W. C. Taylor}
\affiliation{HH Wills Physics Laboratory, Tyndall Avenue, Bristol, BS8 1TL, UK}
\affiliation{Bristol Centre for Functional Nanomaterials, University of Bristol, Tyndall Avenue, Bristol, BS8 1TL, UK}

\author{Malcolm A. Faers}
\affiliation{Bayer AG, 40789, Monheim am Rhein, Germany.}

\author{J. L. Ross Anderson}
\affiliation{School of Biochemistry, University of Bristol, Bristol, BS8 1TD, UK.}
\affiliation{School of Cellular and Molecular Medicine, University Walk, Bristol, BS8 1TD, UK.}
\affiliation{BrisSynBio Synthethic Biology Research Centre, Life Sciences Building, Tyndall Avenue, Bristol, BS8 1TQ, UK.}

\author{Annela M. Seddon}
\affiliation{HH Wills Physics Laboratory, Tyndall Avenue, Bristol, BS8 1TL, UK}
\affiliation{Bristol Centre for Functional Nanomaterials, University of Bristol, Tyndall Avenue, Bristol, BS8 1TL, UK}

\author{C. Patrick Royall}
\email{paddy.royall@espci.fr.}
\affiliation{HH Wills Physics Laboratory, Tyndall Avenue, Bristol, BS8 1TL, UK}
\affiliation{Bristol Centre for Functional Nanomaterials, University of Bristol, Tyndall Avenue, Bristol, BS8 1TL, UK}
\affiliation{School of Chemistry, University of Bristol, Cantock's Close, Bristol, BS8 1TS, UK.}
\affiliation{Gulliver UMR CNRS 7083, ESPCI Paris, Universit\' e PSL, 75005 Paris, France.}

\date{\today}

\begin{abstract}
While proteins have been treated as particles with a spherically symmetric interaction, of course in reality the situation is rather more complex. A simple step towards higher complexity is to treat the proteins as non--spherical particles and that is the approach we pursue here. We investigate the phase behavior of enhanced green fluorescent protein (eGFP) under the addition of a non--adsorbing polymer, polyethylene glycol (PEG). From small angle x-ray scattering we infer that the eGFP undergoes dimerization and we treat the dimers as spherocylinders with aspect ratio  $L/D-1 = 1.05$. Despite the complex nature of the proteins, we find that the phase behaviour is similar to that of hard spherocylinders with ideal polymer depletant, exhibiting aggregation and, in a small region of the phase diagram, crystallization. By comparing our measurements of the onset of aggregation with predictions for hard colloids and ideal polymers [S.V. Savenko and M. Dijkstra, J. Chem. Phys \textbf{124}, 234902 (2006) and F. lo Verso \textit{et al.}, Phys. Rev. E \textbf{73}, 061407 (2006)] we find good agreement, which suggests that the eGFP proteins are consistent with hard spherocylinders and ideal polymer. 
\end{abstract}

\maketitle

\section{Introduction}
\label{sectionIntroduction}

Protein aggregation, and crystallization behavior has important consequences in determining their structure, functions and understanding pivotal challenges ranging from condensation diseases \cite{schneider2012,morris2009,cohen2013} to the development of new materials.\cite{bai2016} 
Controlling their assembly into states in which their functionality can be exploited is crucial to fully realise their potential. Predictions can be made based on phase diagrams, with detailed conditions such as protein concentration, temperature, pH, etc.\cite{alberti2019,lai2012,dumetz2008}

Crystallization is one of the most complex and least understood topics in biology. Protein crystallization is fundamental to obtain protein structure and to get insights into protein function. However, crystallization protocols are mainly based on trial and error assays, with a lack of standardized approaches. In fact, in average only 0.04\% of crystallization experiments yields good quality crystals. \cite{fusco2016} This is due in part to inherent protein shape and surface complexity as well as the dependence of protein-protein interactions on combinations of pH, temperature and precipitants (salts and polymers).\cite{hui2003,huang1999,bhamidi2005,fusco2016,glaser2019,nicolai2013}

Recent developments include an improved understanding of the role of clusters in protein crystallization,\cite{yamazaki2017} the effect of polymers in inducing crystallization,\cite{mcpherson1976, tanaka2002, tardieu2002} the effects of salts on crystallization~\cite{tardieu2002,yamanaka2011,velev1998} and crystal growth rate,\cite{schmit2012} and the role of temperature in the protein 
phase diagram.\cite{astier2008,grouazel2006} An emphasis has been placed on the role of entropy in contact--contact interactions in proteins.\cite{cieslik2009} Perhaps unsurprisingly, fine manipulation of protein interactions is necessary for self-assembly, \cite{kim2015,mandell2009} and if this can be successfully achieved and coupled with protein engineering, it is possible to manipulate proteins to enable new paths of self--assembly.\cite{suzuki2016,zhang2020,bai2016}

By contrast, the field of \emph{Soft Matter} often operates at rather larger lengthscales than the supramolecular lengthscale of proteins. Yet phenomena exhibited by soft materials have been applied to proteins, with some degree of success.\cite{mcmanus2016,fusco2016,stradner2020,hamley2007} Among these is the concept inspired by colloidal systems of effective interactions between the protein molecules that can be altered by other components in the system such as added salts and polymers. \cite{dorsaz2012, zhang2012, zhang2017,piazza2004} In this way a soft matter perspective offers some insights to understand and quantify protein interactions and their equilibrium phase diagrams by simplified models, which might provide a systematic way to improve protein crystallization. \cite{mcmanus2016,fusco2016,stradner2020,dorsaz2012,hamley2007} Indeed, a parameter that has been used to relate protein phase behavior \cite{tessier2002,piazza2004,george1994,elcock2001,wentzel2008} to that of colloidal suspensions \cite{noro2000,lu2008,royall2018} is the second viral coefficient, which can be calculated experimentally from osmotic pressure,\cite{elcock2001} static light scattering, dynamic light scattering, SAXS or SANS\cite{zhang2008,foffi2014,kim2008,huang2019}. While proteins often aggregate at high concentration, some do not and indeed interesting glassy behaviour reminiscent of hard sphere colloids has been seen for concentrated solutions of eye lens $\alpha$-crystallin \cite{foffi2014}, which opens the potential for further analogies with colloidal systems.

Examples of the insights gained from this approach of comparing proteins to colloidal systems include the prediction of enhanced nucleation rates in the vicinity of a (metastable) critical point, \cite{tenwolde1997,pellicane2004} which have been realized using colloids with a short--ranged attraction\cite{savage2009,taylor2012} and gelation\cite{cardinaux2007,kulkarni2003,zhang2008} and so--called liquid--liquid phase separation. \cite{broide1991} It is also possible to control the pathway of crystallization by manipulating the interactions.\cite{whitelam2010} Analogies have also been made between proteins and colloids with so--called mermaid (short--range repulsion, long--range attraction) interactions, through the discovery of finite--sized clusters, \cite{groenewold2001,stradner2004,sedgwick2004,campbell2005,sciortino2005,malins2011,klix2010,vangruijthuijsen2018} although 
the existence of the protein clusters has been questioned. \cite{shukla2008} Meanwhile the colloid systems have been shown to exhibit much more complex behavior than was originally supposed, through a fundamental breakdown in spherical symmetry in the electrostatic repulsions. \cite{klix2013,royall2018} Moreover simple isotropic models applied to other colloidal systems \cite{likos2001,velev1998} often fall short and fail to fully characterize key protein phenomenology due to anisotropic shape and a non-uniform surface charge and hydrophobic/hydrophilic pattern in proteins.\cite{elcock2001,fusco2016,baaden2013} This anisotropy is responsible for the directional and localized protein interactions that yields non-closed packed crystal structures as well as directed self-assembly.\cite{liu2009} This more complex behavior can be captured and reproduced to some extent via \emph{patchy particle} models, where the simulated particles include angular surface directionality of attractive short-range interactions.\cite{glaser2019,fusco2014,doye2006} By changing the number, size and specificity of said patches, the system can be optimised to fully describe the protein behavior. \cite{fusco2014,altan2018,james2015,li2012,gnan2019,cai2018,zhang2020}

In colloidal systems, an effective attraction between the particles can be induced by adding non--adsorbing polymers.\cite{asakura1954,poon2002} This can lead to gelation\cite{poon2002,zaccarelli2007,royall2021,asherie1996} and enhanced crystallization rates.\cite{savage2009,taylor2012} Although they are often smaller than colloids, polymers are typically rather larger than proteins,\cite{vivares2002,zhang2017} leading to the concept of the \emph{protein limit},\cite{bolhuis2003,mutch2007,mutch2008} where the polymers are so much larger than the proteins that the relevant lengthscale is the \emph{intra--}polymer persistence length, rather than the polymer radius of gyration that is typically considered in the case of colloid--polymer mixtures. However, here we consider a scenario more akin to colloid--polymer mixtures, where the polymer radius of gyration $R_g$ is smaller than or comparable to the protein radius.

Polymer--induced protein precipitation has been investigated via volume exclusion interactions i.e. depletion (for high molecular weight polymer). \cite{atha1981,hui2003} However, some polymers can also interact with positively charged amino acids (lysine, arginine and histidine),\cite{hasek2006} and/or through hydrophobic chemical interactions (for example with -CH$_{2}$OCH$_{2}$- groups)\cite{Shkel2015} both present on the protein surface. Additionally, there can be a preferential formation of hydrogen bonds between the polymer and water, which in turn enhances protein-protein interactions.\cite{durbin1996} These scenarios can lead to more complex interactions than the non-adsorbing polymer-protein depletion picture.

Here, we consider mixtures of \emph{enhanced} Green Fluorescent Protein (eGFP) and poly--ethylene--glycol (PEG). eGFP readily undergoes dimerization\cite{kim2015} such that the proteins resemble short rods. By comparing our results to the extensive colloid--polymer literature,\cite{poon2002} we treat the proteins as spherocylinders (with dimensions deduced from x--ray scattering), in particular as mixtures of hard spherocylinders and ideal polymer \cite{bolhuis1997,savenko2006}. In this way, a we consider a model of the protein--polymer mixture where the only level at which the complexity of the system is treated is via a simplified form for the anisotropy of the protein dimers, namely a spherocylinder. We thus treat the protein dimers as hard spherocylinders and the polymers as ideal polymers.

Here we follow the literature on spherocylinder--polymer mixtures \cite{bolhuis1997,savenko2006} and express the aspect ratio as $L/D-1$ in which the aspect ratio of spheres is then zero where $L$ is the spherocylinder length and $D$ is the diameter. We interpolate the predictions of polymer \emph{fugacity} required for polymer--induced demixing, between spherocylinders with aspect ratio $L/D-1 = 5$\cite{savenko2006} and spheres \cite{loverso2006}. Remarkably, given the simplicity of the model, we find good agreement for the geometric parameters of our system with $L/D-1 = 1.05$.

This paper is organised as follows. In section \ref{sectionMethodology} we describe the methods of protein preparation, characterization, as well as estimation of their interactions as spherocylinders. Section \ref{sectionPhase} consists of phase behavior, including aggregation and crystallization, in salt-screened and salt-free mixtures. We then compare the onset of aggregation with theory in Section \ref{sectionComparison}, where the polymer radius of gyration at different molecular weights are fitted by interpolating predictions from hard colloids and ideal polymers. Finally, a discussion of our findings presented in Section \ref{sectionDiscussion}.

\section{Methodology}
\label{sectionMethodology}

\subsection{Estimation of Interactions Between Proteins}
\label{sectionEstimationInteraction}

Our system is governed by two control parameters: the protein concentration and the polymer concentration. In the context of treating the system in the spirit of a colloid--polymer mixture, we consider the proteins as spherocylinders and thus the volume fraction

\begin{equation}
\phi_\mathrm{eGFP}=\rho_{\mathrm{eGFP}}\left(\frac{\pi}{6} D^{3}+\frac{\pi}{4} D^{2} (L-D)\right)
\label{eqPhi}
\end{equation}

\noindent where $\rho_\mathrm{eGFP}$ is the protein number density. We determine the diameter $D$ and length $L$ from x--ray scattering and compare our results to literature values (see section \ref{sectionCharacterisation}). Our choice of spherocylinders is motivated by the literature on colloid-polymer mixtures, for which phase diagrams for hard spherocylinders plus ideal polymer are available.\cite{bolhuis1997,savenko2006,lekkerkerker2011}

Our second control parameter, polymer concentration, is expressed as the polymer fugacity $z_\mathrm{pol}$.  We make the significant assumption that the polymers can be treated as an ideal gas (of polymers), and then the fugacity is equal to the polymer number density in a reservoir $z_\mathrm{pol}=\rho_\mathrm{pol}^\mathrm{res}$ in thermodynamic equilibrium with the system.

To compare with predictions from theory and computer simulation, we use the fraction of available volume $\alpha$ to estimate the reservoir polymer number density from that in the experimental system $\rho_p^\mathrm{exp}$, viz $\rho_p^\mathrm{exp} = \alpha \rho_p^\mathrm{res}$. We use the free-volume approximation for $\alpha$.\cite{lekkerkerker2011}

\begin{widetext}
\begin{equation}
\begin{array}{c}
\alpha=\left(1-\phi_{\mathrm{eGFP}}\right) \exp \left[-\left\{A\left(\frac{\phi_{\mathrm{eGFP}}}{1-\phi_{\mathrm{eGFP}}}\right)+B\left(\frac{\phi_{\mathrm{eGFP}}}{1-\phi_{\mathrm{eGFP}}}\right)^{2}+C\left(\frac{\phi_{\mathrm{eGFP}}}{1-\phi_{\mathrm{eGFP}}}\right)^{3}\right\}\right] \vspace{2ex}\\
A=\frac{6 \gamma}{3 \gamma-1} q+\frac{3(\gamma+1)}{3 \gamma-1} q^{2}+\frac{2}{3 \gamma-1} q^{3} \vspace{2ex}\\
B=\frac{1}{2}\left(\frac{6 \gamma}{3 \gamma-1}\right)^{2} q^{2}+\left(\frac{6}{3 \gamma-1}+\frac{6(\gamma-1)^{2}}{3 \gamma-1} \right) q^{3} \vspace{2ex}\\
C=\frac{2}{3 \gamma-1}\left(\frac{12 \gamma(2 \gamma-1)}{(3 \gamma-1)^{2}}+\frac{12 \gamma(\gamma-1)^{2}}{(3 \gamma-1)^{2}} \right) q^{3}
\end{array}
\label{eqfreevolume}
\end{equation}
\end{widetext}

\noindent
where $\gamma = L/D$ is the length-to-diameter ratio of spherocylinders, and $q$ is the polymer-protein size ratio, of $2R_g/D$, $R_g$ is the radius of gyration of polymers. Below we compare the phase behavior we obtain for our system with literature values for spherocylinder--polymer mixtures.\cite{bolhuis1997,savenko2006,lekkerkerker2011}

Our proteins carry an electrostatic charge, which we determine below (section \ref{sectionCharacterisation}). To estimate the electrostatic interactions we used a screened Coulomb (Yukawa) potential. Here, it is convenient to treat the proteins as spheres. We shall see below that although they are not spherical, the electrostatic interactions turn out to be so weak that we believe that to a large extent they can be neglected. Therefore we merely estimate their strength with a spherically--symmetric approximation.

\begin{equation}
\beta u_\mathrm{yuk}(r)=\left\{\begin{array}{ll}
\infty & \text { for } r< D \\
\beta \epsilon_\mathrm{yuk} \frac{\exp (-\kappa(r-D))}{r / D} & \text { for } r \geq D
\end{array}\right.
\label{eqYukawa}
\end{equation}

\noindent 
where the contact potential,

\begin{equation}
\beta \epsilon_\mathrm{yuk}=\frac{\mathscr{Z}^2}{(1+\kappa D / 2)^{2}} \frac{\lambda_B}{D}
\label{eqEpsilon}
\end{equation}

\noindent 
$\kappa$ is the inverse Debye length,

\begin{equation}
\kappa^{2}=4 \pi \lambda_{\mathrm{B}} \sum_{i} \rho^\mathrm{ion}_{\mathrm{i}} (Z_{i}^\mathrm{ion})^{2}
\label{eqKappa}
\end{equation}

\noindent
with $\mathscr{Z}$ the charge number, $\lambda_{\mathrm{B}}$ is the Bjerrum length, $\rho_i^\mathrm{ion}$ is the number density of the $i$th ionic species, $\mathscr{Z}_i^\mathrm{ion}$ is the valency of the $i$th ionic species. For the system with no added salt, the ionic strength $I = \sum_{i}\rho^\mathrm{ion}_\mathrm{i} (\mathscr{Z}_{i}^\mathrm{ion})^{2}$ was evaluated as the sum of the ion contributions of the weak dissociation of 25 mM HEPES (pKa = 7.66) and the counter ions contribution assuming charge neutrality. Thus, by varying protein concentrations we obtained a range of $I$ = 1--4 mM. For the system where 10 mM NaCl was added, we included to the sum the ions contribution from this salt dissociation, giving a range of $I$ = 15--18 mM. Further details of Yukawa potential are listed in Table \ref{tableParameters}.

It is important to highlight, as pointed out by Roosen-Runge and collaborators,\cite{roosen2013} that in addition to assuming a spherical shape for the proteins, also an isotropic distribution of ions on their surfaces is assumed. This is not the case for eGFP dimers, thus the charges calculated should only be considered as effective charges suitable to describe the phenomena observed in our experiments. However, the magnitude of the charge that we determine is sufficiently small that within the DLVO treatment we employ, the electrostatic interactions prove to be very weak, so we believe that at the level of this analysis, a spherical approximation is reasonable.

\subsection{Protein Expression and Purification}
\label{sectionProteinExpression}

\textbf{Cellular Culture for the Expression of eGFP.} A mini-culture of competent \emph{Escherichia coli} BL21 (DE3) previously transformed with the DNA plasmid-pET45b(+)-eGFP was prepared by inoculating 100 mL of lysogeny broth (LB) and the antibiotic carbenicillin (50 $\mu$g/mL) with an isolated \emph{E. coli.} colony. The culture was left to grow overnight (16 h) at 37$^{\circ}$ C and 180 rpm. 2 mL of this culture inoculated to a 1 L of LB containing the same antibiotic and which was left to grow under the same previous conditions. The optical density (OD$_\mathrm{600nm}$) was monitored until a value of 0.5--0.6 was reached. Then, the production of eGFP was induced by adding 1 mM of Isopropyl $\beta$-D-1-thiogalactopyranoside (IPTG). After 1 h of induction time, the temperature was changed to 28$^{\circ}$ C and was incubated overnight. The cell culture was centrifuged at 4500 g for 15 min at 4$^{\circ}$ C. The supernantant obtained was discarded and the pellet was resuspended in a \emph{lysis buffer} (20 mM imidazole, 300 mM NaCl and 50 mM potassium phosphate at pH 8.0) and stored at -20$^{\circ}$ C.\cite{tang2018}

\textbf{Purification and concentration of eGFP.} Cell pellets were thawed and kept on ice, sonicated for 3 cycles of 30 seconds (Soniprep 150 plus MSE) and centrifuged at 18000 rpm (Sorvall SS34 rotor) at 4$^{\circ}$ C for 30 min. The supernatant was recovered and filtered through a 0.22 $\mu$m syringe filter (Millipore) and injected to a Ni-NTA (nickel-nitrilotriacetic acid) agarose column (Qiagen) connected to an \"AKTA START purification system (GE Healthcare), previously equilibrated with the lysis buffer mentioned. The bound eGFP was washed with the same lysis buffer to elute the rest of unbound proteins and eGFP was later eluted with a linear gradient (0--100\%) of a 500 mM imidazole, 300 mM NaCl, 50 mM potassium phosphate buffer at pH 8.

The recovered proteins were then further purified through size exclusion chromatography to eliminate aggregates and unfolded protein. A single peak corresponding to a single protein size was collected. eGFP was concentrated to $\sim$3 mL in a 25 mM Tris-Base 150 mM NaCl buffer at pH 7.4. The proteins were applied to a HiLoad Superdex 75 16/600 size exclusion column using  \"AKTA START purification system (GE Healthcare) pre-equilibrated with the same buffer. Protein elution was monitored at 280 nm.

Purified eGFP was filtered through a 0.22 $\mu$m syringe filter (Millipore) and concentrated using protein 30 kDa concentrators (ThermoFisher Scientific) at 5000 rpm and 4$^{\circ}$ C for the time required to reach the desired volume. The protein concentration was determined by measuring the absorbance at $\lambda_\mathrm{eGFP}$ = 488 nm with a molar extinction coefficient $\epsilon_\mathrm{eGFP}$ = 56000 M$^{-1}$cm$^{-1}$. \cite{kaishima2016}

\begin{figure*}[ht!]
\includegraphics[height=0.3\textwidth]{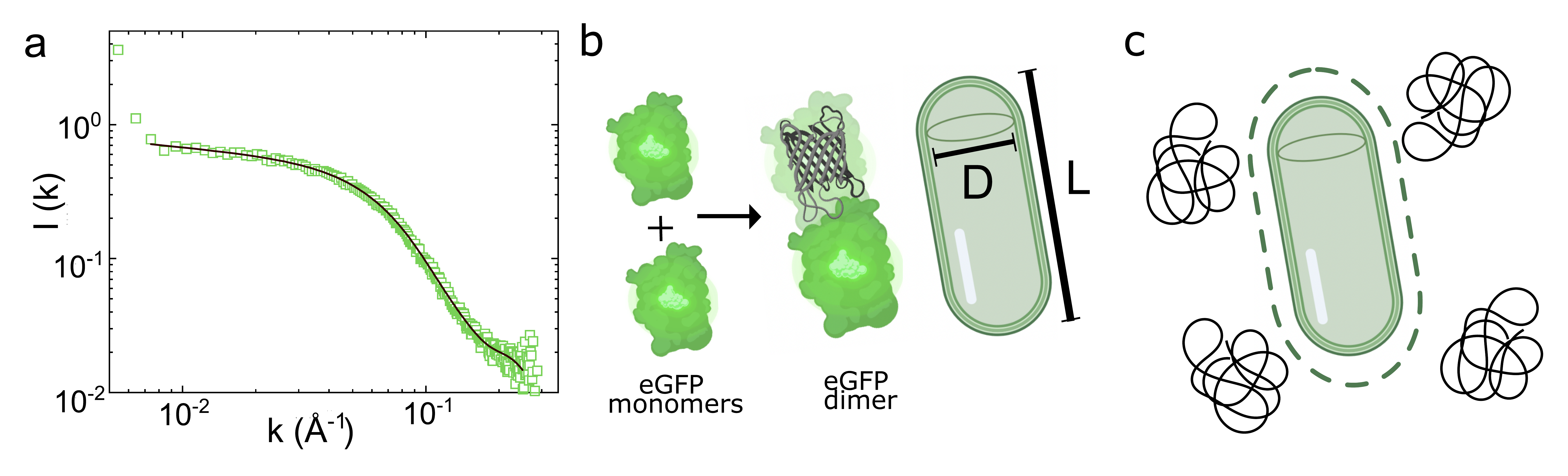}
\caption{\label{figSAXS} Determining protein form factor with small angle X-ray scattering (SAXS). (a) SAXS scattering intensity with a cylinder model fitting (line) for 10 mg/mL protein solutions of eGFP. (b) Illustration of the dimer structure of eGFP and the spherocylinder representation used for data analysis, showing the diameter $D$ and the length $L$. (c) Schematic of eGPF depletion zone when polymers are present.
}
\end{figure*}

\textbf{Sample preparation}. 
From small-angle X-ray scattering, the purified eGFP showed to form dimers with a height of 8.2 nm and length of 4 nm, as shown in Fig. \ref{figSAXS}. Thus we treated the eGFP molecules as spherocylinders, which are made of two hemispheres of diameter $D = 4$ nm, length $L=8.2$ nm. We changed the protein buffer to 25 mM HEPES at pH = 7.4. A separate buffer solution with 0 or 200 mM NaCl was used as a stock solution to adjust the final protein and salt concentrations.

We carried out experiments with two different polymer sizes, in particular, polyethylene glycol (PEG, Polymer Laboratories) with molecular weights $M_w$ of 620 and 2000. The polymer radius of gyration $R_g$ was estimated from polymer scaling with an empirical prefactor $R_g^{\mathrm{scale}}=0.020 M_{\mathrm{w}}^{0.58}$,\cite{crys1994} leading to $R_g^\mathrm{scale}$ of 0.83 and 1.64 nm, and a polymer-protein size ratio ($q^\mathrm{scale}=2R^\mathrm{scale}_g/D$) of 0.42 and 0.82, for the small and large polymers respectively.

For each sample, we first mixed different volumes of the protein stock solution (106.4 mg/mL) with different volumes of the HEPES buffer with and without salt to complete a fixed volume of 5 $\mu$L, giving a range of protein concentrations of 0.7--30 mg/mL and a constant NaCl concentration of 10 mM (for the samples with salt). To induce effective attractions between the protein molecules, we added different amounts of PEG by weight at room temperature such that we obtained a polymer concentration between 0--0.8 gcm$^{-3}$ (fugacity $\sim 1$--$50$). The samples were thoroughly shaken for 5 seconds by a touch-vortexer, immediatly imbibed to inside-diameter of 0.5 mm capillaries (CM Science), and sealed with optical adhesive (Norland Optical no 81). Within 5 minutes, the different phases obtained were characterised through laser scanning confocal microscopy (Leica SP8) at an excitation wavelength of 488 nm and emission wavelength of 509 nm.

\vspace{20pt}
\subsection{Characterisation}
\label{sectionCharacterisation}

\vspace{10pt}
\noindent\textbf{SAXS analysis}
To characterise the size and shape of the expressed and purified eGFP, we performed SAXS measurements on 25 $\mu$L of 10 mg/mL eGFP in 25 mM Tris-Base 150 mM NaCl buffer at pH 7.4 on a SAXSLAB Ganesha 300XL instrument. Samples were loaded into 1.5 mm borosilicate glass capillaries (Capillary Tube Supplies UK) and sealed with optical adhesive under UV light (Norland 81). The wavevector $k$ range was of 0.006--0.30 $\AA$$^{-1}$. Background corrections were carried out with both an empty cell and a cell with the buffer only. The obtained data were fitted using the SasView version 4.0 software package.\cite{sasview}

The results are shown in Fig. \ref{figSAXS}\textbf{a}. The scattering intensity, \emph{I(k)} is given by the product of the form factor $P(k)$ and the static structure factor $S(k)$ via

\begin{equation}
I(k) = \phi_{\mathrm{eGFP}}  V_\mathrm{eGFP}(\Delta\rho_\mathrm{scat})^{2}P(k)S(k)
\label{eqI}
\end{equation}

\noindent
$V_\mathrm{eGFP}$ is the volume of a protein dimer (Eq. \ref{eqPhi}) and $\Delta \rho_\mathrm{scat}$ is the difference in scattering length density between the proteins and its solvent.\cite{zhang2007,wolf2014,singh2019} The scattering data was successfully fitted by a cylindrical form factor\cite{arpino2012} with a diameter of 4.0 $\pm$ 0.02  nm and a length of 8.2 $\pm$ 0.08 nm (see full parameters in Appendix A \ref{tableSAXSfitting}). These dimensions are consistent with dimers of eGFP as illustrated in Fig. \ref{figSAXS}\textbf{b}. These results are in agreement with previous work on eGFP, where it was found that the protein exists in dimers.\cite{myatt2017}

\textbf{Electrophoretic Mobility Measurement.} We performed electrophoretic mobility $\mu_{e}$ measurements on 1 mL protein solutions of 2 mg/mL at 20$^{\circ}$ C in a NaCl 10 mM solution using a Zetasizer Nano ZS (Malvern, UK) at a detector angle of 13$^{\circ}$ and a 4 mW 633 nm laser beam to determine the charge of eGFP following Roosen-Runge, \emph{et al}.\cite{roosen2013} Care was taken in order to have the same pH with the buffer used in phase diagram determination. By using electrophoretic light scattering (ELS) via phase analysis light scattering (M3-PALS), the electrophoretic mobility $\mu_{e}$ of eGFP was determined as an external electric field is applied.

From this we obtained the zeta potential $\zeta$ for a spherical particle with diameter \emph{D} using

\begin{equation}
\mu_e=\dfrac{2 \epsilon \zeta f(\kappa D/2)}{3 \eta}
\label{eqMu}
\end{equation}

\noindent
where $\epsilon_r$ and $\eta$ are the dielectric constant and the viscosity of the solvent, respectively, and $f(\kappa D/2)$ is the Henry function evaluated at $\kappa D/2$. The relation between surface charge density $\sigma$ and the reduced zeta potential \emph{$\tilde{\zeta}$ = ($e\zeta$)/(2k$_{B}$T)} is:

\begin{widetext}
\begin{equation}
\frac{\sigma e}{k_{B}T}=2\epsilon_r \kappa \left [ \sinh^{2}\left ( \frac{\tilde{\zeta}}{2} \right ) + \frac{4}{\kappa D} \tanh^{2} \left ( \frac{\tilde{\zeta}}{4} \right ) + \frac{32}{(\kappa D)^{2}} \ln \left ( \cosh \left ( \frac{\tilde{\zeta}}{4} \right ) \right )  \right ]^{1/2}
\label{eqZetaPotential}
\end{equation}
\end{widetext}

\noindent
Finally we can obtain the total charge using $\mathscr{Z}e=\pi D^2\sigma$, where $\mathscr{Z}$ is the charge number and $e$ is the elementary charge. The zeta potential value measured was $\zeta$ = -7.02 mV, which corresponds to a charge number of $\mathscr{Z}$ = 1.16. We list the parameters for Yukawa potential in Table \ref{tableParameters}.

\section{Results}
\label{sectionResults}

We divide our results as follows. First a phase diagram is presented for the eGFP-PEG620 system, showing the fluid-aggregation  transition in section \ref{sectionPhase}. We increased the polymer molecular weight to obtain a larger size ratio (using PEG2000), investigating the effects of polymer size on phase boundary. To check any residual effects of protein charges, the comparison between salt-screened and salt-free system is discussed. In section \ref{sectionComparison}, we consider a spherocylinder-sphere system of $L/D-1 = 1.05$. The polymer radius of gyration is fitted by interpolating between theoretical and computer simulation predictions. Finally the protein crystallization, formed through depletion attractions with polymer, is discussed in section \ref{sectionDiscussion}.

\subsection{Phase Behavior}
\label{sectionPhase}

\begin{figure*}[t!]
\centering
\includegraphics[width=18cm]{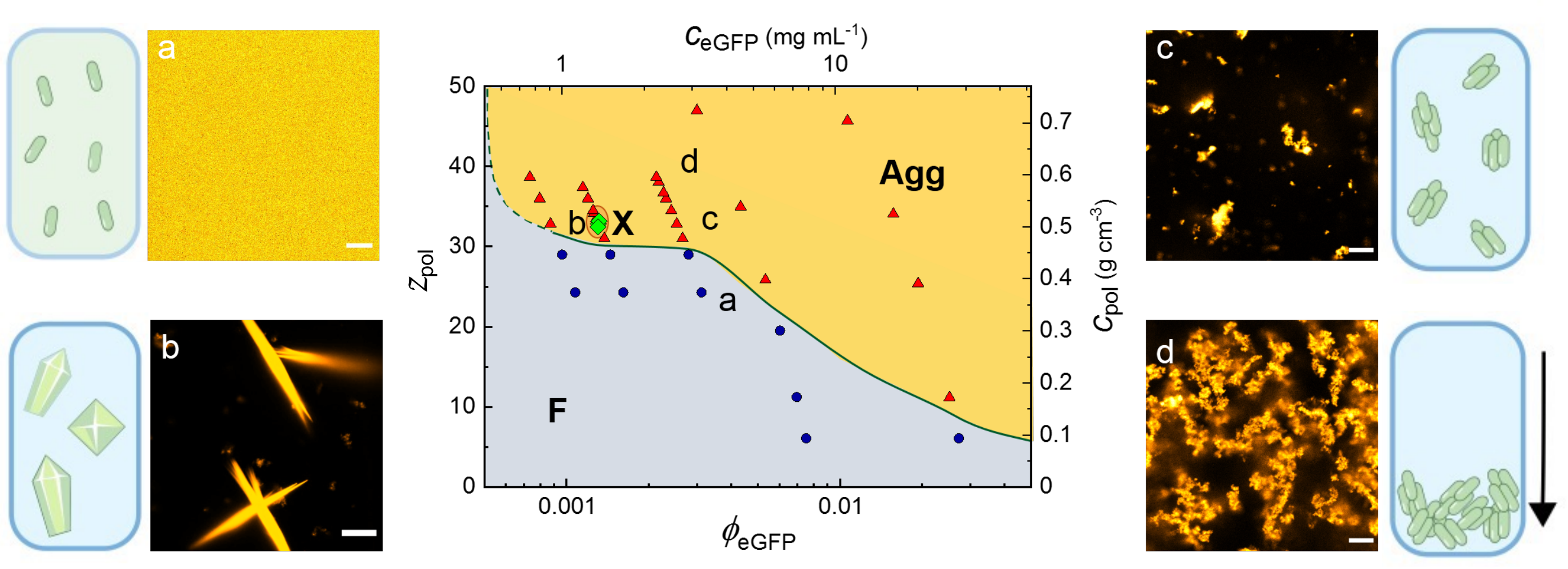}
\caption{Phase diagram of the eGFP -- PEG system with 10 mM NaCl. Here the phase behavior is shown in the protein volume fraction -- polymer fugacity ($\phi_\mathrm{eGFP}-z_\mathrm{pol}$) plane, coupled with their concentrations. Fluids, aggregating systems and crystals are denoted by blue circles, red triangles, and green diamonds respectively. 
Confocal microscopy images of different states, along with a schematic representation of the system behavior are shown in (a)-(d) as follows: 
(a) fluids, 
(b) crystals, 
(c) aggregation, 
(d) denser sediments. 
 The scale bars indicate 10 $\mu$m.  
\label{figSalted}}
\end{figure*}

\textbf{Salt--screened system.}
The phase diagram with different states as determined from images from confocal microscopy for the eGFP -- PEG620 (small polymer) system with 10 mM of added NaCl salt is shown in Fig. \ref{figSalted}. The phase diagram is presented in the plane of protein volume fraction ($\phi_\mathrm{eGFP}$) and polymer fugacity $z_\mathrm{pol}$. The phase boundaries are determined by the average between the fluid and aggregated phase points. Note that in depletion systems, aggregation and gelation are identified with the liquid--gas phase boundary. \cite{lu2008,royall2018,royall2021} Thus while these are non--equilibrium states, comparison with equilibrium phase behaviour is nevertheless highly informative. For lower protein volume fraction below we tested, a dotted line is drawn based on the intuition from literature.\cite{poon2002} The smaller the concentration of proteins, the higher the polymer concentration needed for phase separation. As noted above, the protein volume fraction is estimated by assuming that the eGFP molecules are spherocylinders of aspect ratio $L/D-1 = 1.05$. The polymer fugacity is obtained from the polymer number density. The protein dimensions determined from SAXS (section \ref{sectionCharacterisation}) and estimated polymer size gave a size ratio $q = (2R_g^\mathrm{scale})/D \sim 0.4$.

As a function of polymer concentration, we first encounter protein solutions where the eGFP appears stable and exhibits no observable aggregation, but instead there is a uniform fluorescent intensity, as the protein dimers are far below the resolution of the microscope (Fig. \ref{figSalted}\textbf{a}). Upon increasing the polymer concentration, we see aggregation for polymer fugacity $z_\mathrm{pol} = 30.0 \pm1.0$ (which corresponds to a protein volume fraction around 0.002), shown in Fig. \ref{figSalted}\textbf{c}. Now the polymer concentration here is rather high, indeed the polymer volume fraction $\phi_\mathrm{pol}= \rho_\mathrm{pol}\pi R_g^3/6$ is of order unity. We return to this point below in section \ref{sectionDiscussion}.

As the protein concentration is increased, the polymer concentration required for aggregation decreases. Upon further increase in polymer concentration, protein aggregates form quickly and become large enough that considerable quantities sediment to the bottom of the sample where a denser sediment builds up (Fig. \ref{figSalted}\textbf{d}). This is reminiscent of aggregation and sedimentation behavior in colloidal systems.\cite{piazza2014} In a small region of the phase diagram, we encounter protein crystallization, indicated as green diamonds (see region denoted as ``X'') in Fig. \ref{figSalted}. We note that there is some lack of smoothness in the phase boundary. Such fluctuations in phase boundaries we well--known in soft matter systems (see. e.g. \cite{poon2012}) and we leave this for further investigation.

Protein crystallization has been related to near--critical behavior.\cite{tenwolde1997} Here, although the regime of crystallization occurs near the aggregation line (which, by itself might link it to criticality \cite{tenwolde1996,taylor2012}) the protein volume fraction is vastly lower than any critical isochore that would be expected to occur for this system. Indeed, the volume fraction of the critical isochore for spherical colloids plus polymers with size ratio $q\sim0.4$ is estimated to be at least $\phi_c \gtrsim 0.25$,\cite{loverso2006} so it is hard to imagine that critical fluctuations are important here. The lengths of the crystallites that we find are in the range of $\sim$4--80 $\mu m$. Fig. \ref{figCrystals}\textbf{b} is pure crystal, while Figs. \ref{figCrystals}\textbf{a} and \textbf{c} show aggregates which we presume to be amorphous.

\begin{figure}[ht!]
	\includegraphics[width=9cm]{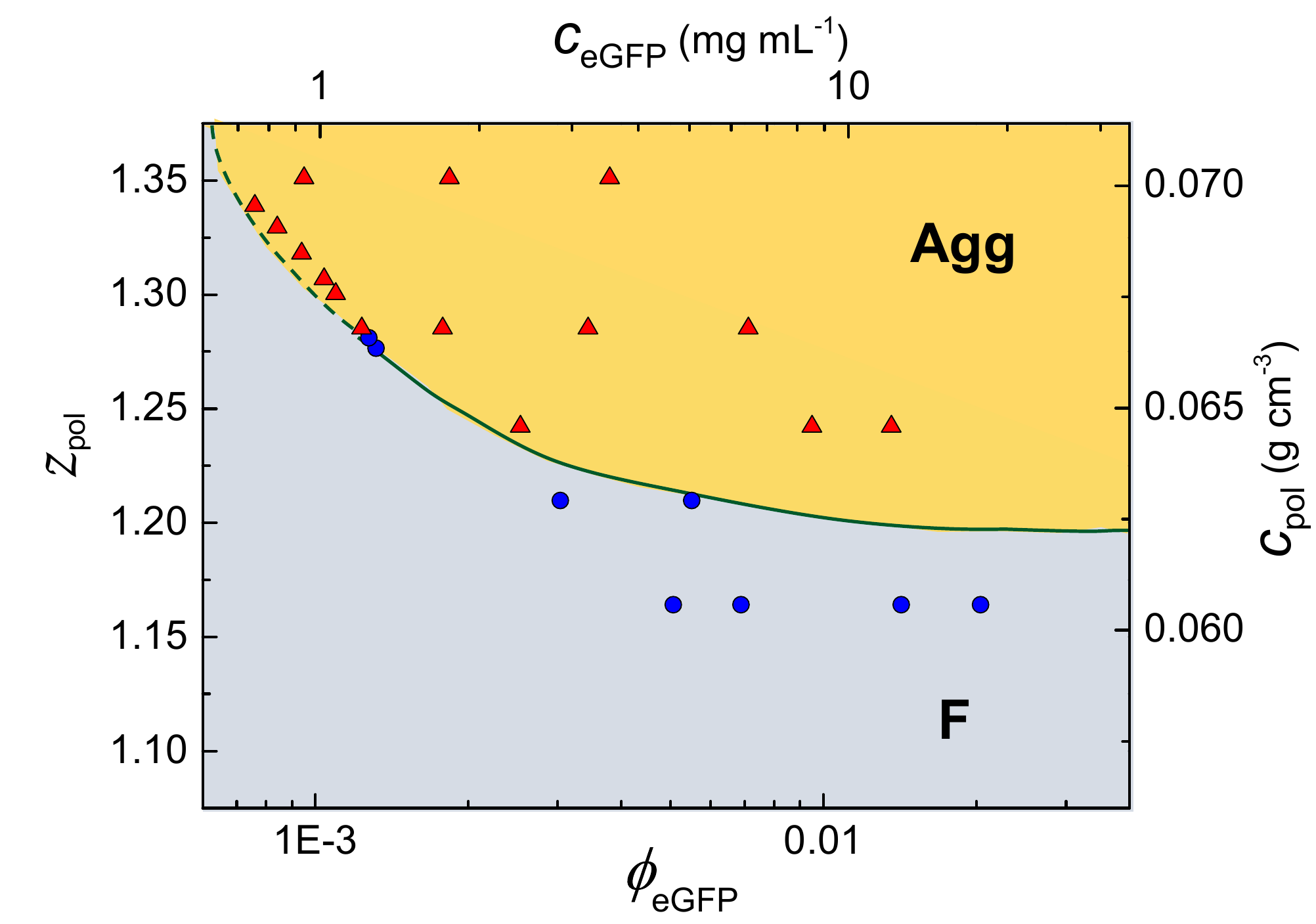}
	\caption{ 
		A Phase diagram of eGFP with PEG2000 and 10 mM NaCl. Fluids and aggregating systems are denoted by blue circles and red triangles, respectively.
		\label{fig2000}}
\end{figure}

\begin{figure}[ht!]
\includegraphics[width=8.5cm]{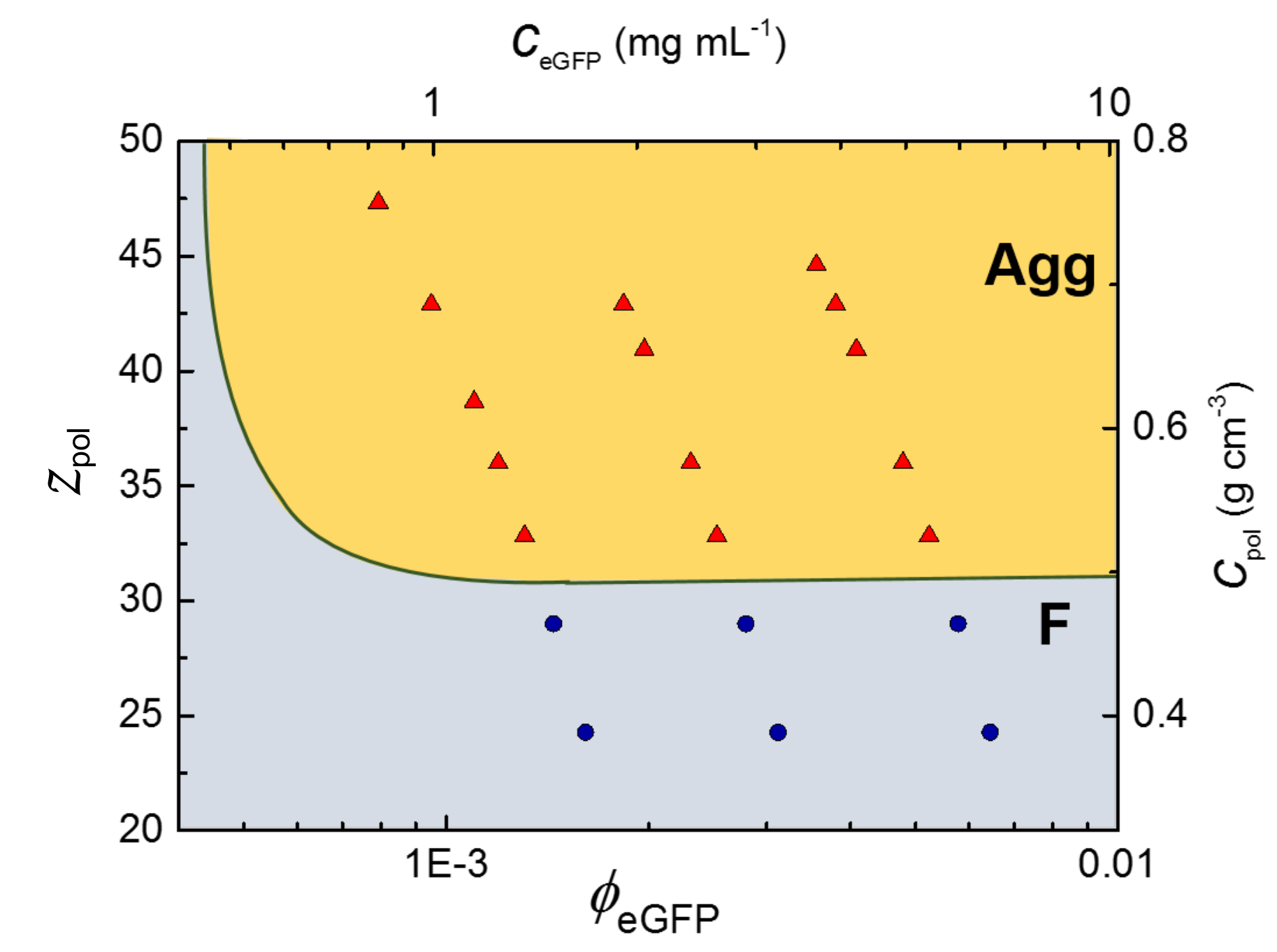}
\caption{Phase diagram of the eGFP -- PEG620 system with no added salt. Fluids and aggregating systems are denoted by blue circles and red triangles, respectively.
\label{figNoSalt}}
\end{figure}

\textbf{Salt--free system.}
To investigate the effect of the (weak) electrostatic interactions between the proteins, we determine the phase behavior in the absence of added salt shown in Fig. \ref{figNoSalt}. We find a boundary for aggregation estimated at $z_\mathrm{pol}=30.9 \pm1.9$ for a rather lower volume fraction of 0.002, which is almost indistinguishable behavior to the case with added salt (Fig.  \ref{figNoSalt}) at the same protein volume fraction. This is quite consistent with the soft matter inspired analysis of treating the proteins as hard spherocylinders. However we do not encounter any crystallization behavior here and return to this in the discussion below.

\textbf{Effects of PEG molecular weight.} 
So far, we have discussed the system with the smaller polymer (PEG620), we now switch to the larger polymer. We chose PEG2000 here because its size is comparable to that of the protein. We therefore expect normal depletion behaviour, as described by the Asakura--Oosawa model, unlike the protein limit $q \gg 1$.\cite{bolhuis2003,mutch2007,mutch2008} 
The phase diagram for the eGFP--PEG2000 system is shown in Fig. \ref{fig2000}. The aggregation shows at a much lower fugacity, $z_\mathrm{pol} = 1.20 \pm0.04$, compared with the smaller polymer at same protein volume fraction of 0.02. This is qualitatively consistent with the literature,\cite{poon2002, lekkerkerker1992, atha1981} that the larger the polymer, the lower fugacity is
needed for phase separation. Below we provide a more quantitative comparison.

\begin{figure*}[ht!]
\includegraphics[width=14cm]{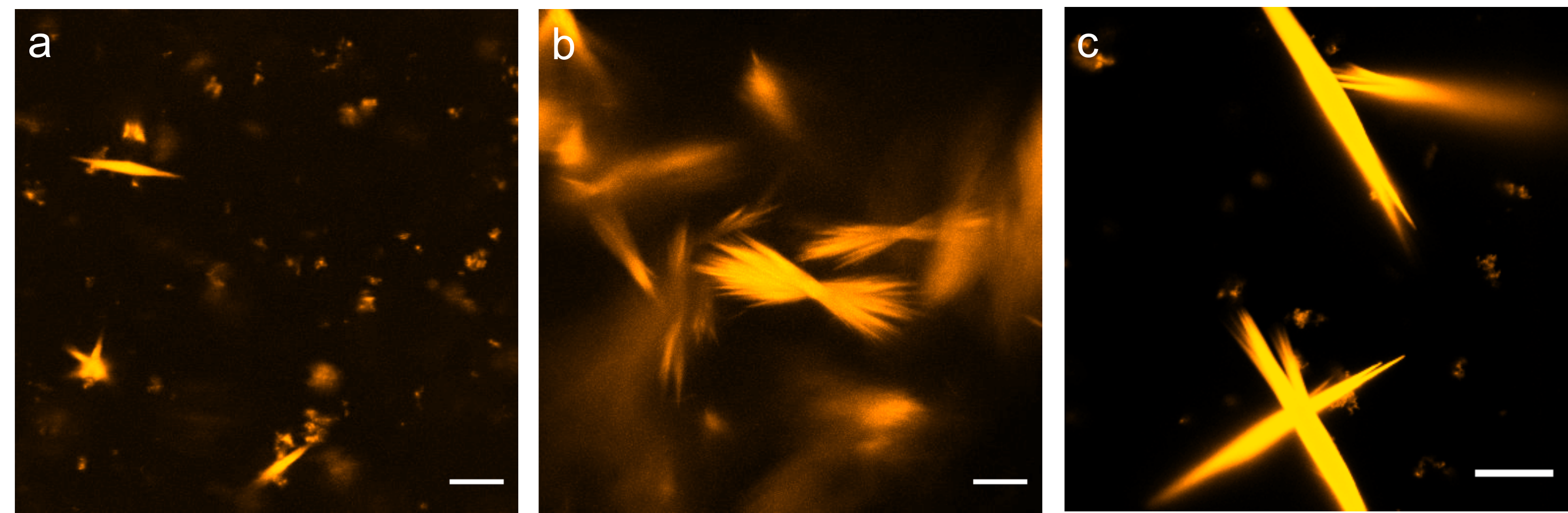}
\caption{\label{figCrystals} Crystallization close to the demixing phase boundary, see Fig. \ref{figSalted}\textbf{b} for location on the phase diagram. Upon increasing the polymer volume fraction, crystals were found around $z_\mathrm{pol}=32.5$ and $\phi_\mathrm{eGFP}$ = 0.0011. Bars = 10 $\mu$m.}
\end{figure*}

\begin{table}
\begin{center}
\caption{\label{tableParameters}Effects of adding salts on Yukawa potential}
\begin{ruledtabular}
\begin{tabular}{ccccc}
NaCl Concentration (mM) & $\kappa^{-1}$ (nm)&$\kappa D$& $Z_e$ & $\beta \epsilon_\mathrm{yuk}$\\
\hline
0 & 4.65 & 1.51& 0.95& 0.0320\\
10& 2.48 &2.82& 1.31& 0.0322\\
\end{tabular}
\end{ruledtabular}
\end{center}
\end{table}

\begin{figure}[ht!]
\centering
\includegraphics[width=8.8cm]{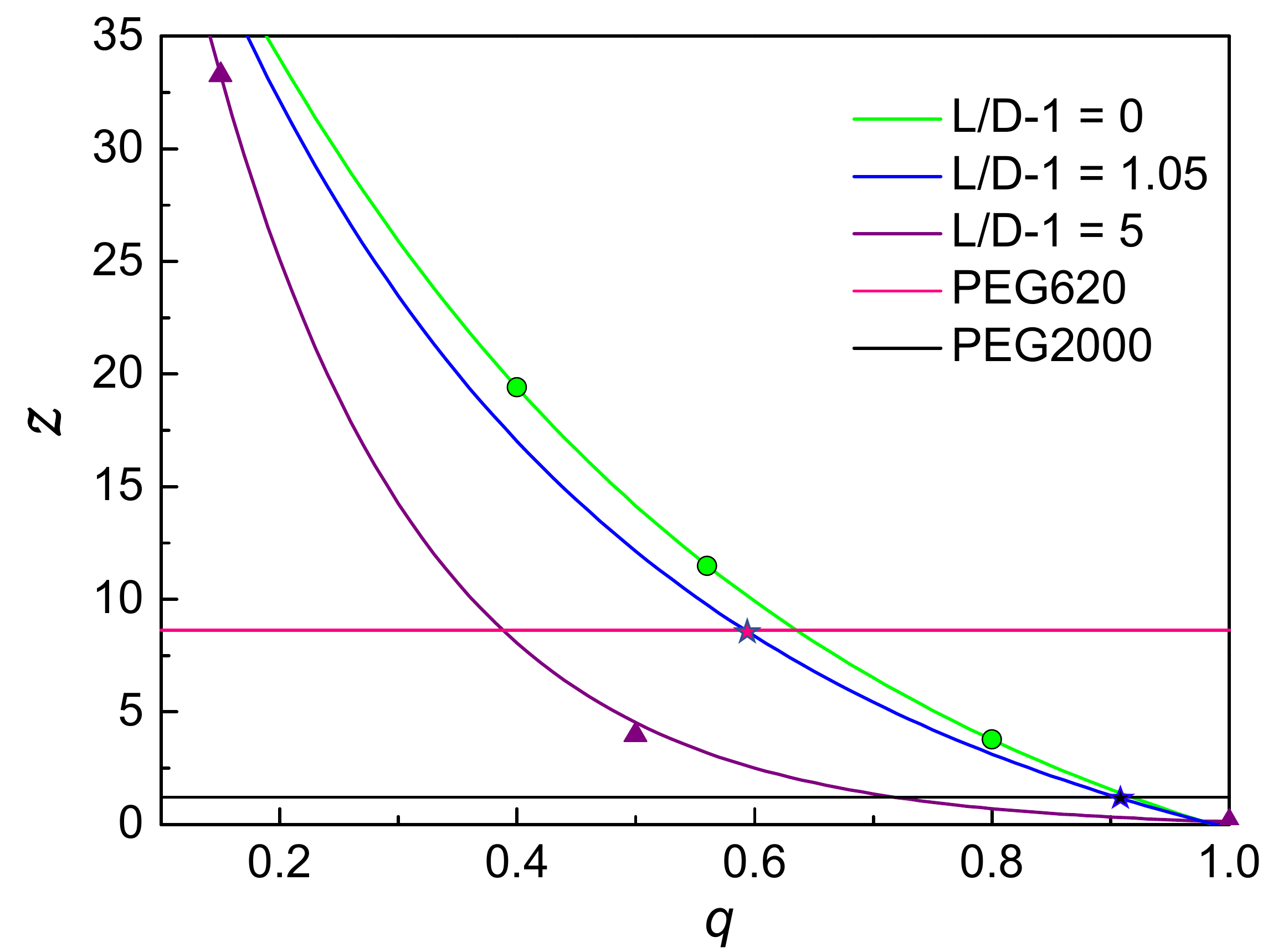}
\caption{\label{figRgFitting} The interpolation from the data presented by Savenko and Dijkstra \cite{savenko2006} for $L/D-1 = 5$ (purple triangles) and Lo Verso et al.\cite{loverso2006} for $L/D-1 = 0$ (green circles), green and purple lines are fitted by power functions. The blue line is the interpolation, and the pink and black stars are crossovers denote matching for PEG620 and PEG2000, respectively.}
\end{figure}

\subsection{Comparison with theory}
\label{sectionComparison}

In order to make a comparison with theoretical and computer simulation predictions, we interpolate between phase boundaries determined for spheres $(L/D-1 = 0)$\cite{loverso2006} and spherocylinders of a larger aspect ratio than those we consider here $(L/D-1 = 5)$ (Fig. \ref{figRgFitting}).\cite{savenko2006} It is important to note what the exact phase \emph{is}. In the case of sphere--polymer mixtures, upon adding polymer at low colloid volume fraction, the first phase transition that is encountered (for $q\gtrsim0.3$) is the (colloidal) liquid--gas demixing.\cite{lekkerkerker1992,poon2002,dijkstra1999,lekkerkerker2011} In the case of spherocylinders with aspect ratio $L/D-1=5$ it is fluid--crystal coexistence.\cite{bolhuis1997,savenko2006} Nevertheless, for spheres at $q\approx0.4-0.5$ the liquid--gas and fluid--crystal phase boundaries occur at quite similar values of the polymer fugacity and so here we neglect the difference. We are, in any case unaware of any computation of the phase diagram for our parameters, and note that the free volume theory of Lekkerkeker \emph{et al.}\cite{lekkerkerker1992} is not highly accurate for these parameters.\cite{loverso2006}

We fit the data for spheres\cite{loverso2006} and spherocylinders\cite{savenko2006} by a power law $z_\mathrm{pol} = a-bq^c$ at a low value of protein volume fraction $\phi_\mathrm{eGFP}\sim0.02$ with different $q$. Here $a$, $b$ and $c$ are fitted constants. The interpolation is done linearly, by $z_\mathrm{pol}^{(0)}-1.05(z_\mathrm{pol}^{(0)}-z_\mathrm{pol}^{(5)})/5$, where $z_\mathrm{pol}^{(0)}$ is for spheres, \cite{loverso2006} and $z_\mathrm{pol}^{(5)}$ is for spherocylinders ($L/D-1 = 5$). \cite{savenko2006} Our interpolation is shown in Fig. \ref{figRgFitting} where we plot the fitted phase boundaries for fitting data and our interpolation. We interpolate to obtain values of $q$ that are consistent with our measured fugacity for demixing $z_\mathrm{pol} = 8.63$ (smaller polymer) and $z_\mathrm{pol} = 1.20$ (larger polymer). We have in addition some uncertainty in determining the size ratio $q$. As noted above, our estimate for the polymer radius of gyration $R_g^\mathrm{scale}$ relied on polymer scaling, which may not be accurate for such small polymers. Moreover there are a variety of other assumptions, such as polymer ideality, rigidity, which have been addressed in more refined theoretical treatments.\cite{lekkerkerker2011,fleer2008} We therefore accept some adjustment in our fitted values and take $q^\mathrm{fit}=0.59$ for smaller polymer and $q^\mathrm{fit}=0.90$ for larger polymer, which agree well with our data.

For larger polymers the fitted polymer radius of gyration $R_g^\mathrm{fit}$ of 1.80 nm falls close to the one from empirical equation of $R_g^\mathrm{scale}=1.64$ nm (see section \ref{sectionProteinExpression}). It is worth noting that there is an fitted increase from the size ratio from scaling $q^\mathrm{scale}=0.42$ to the fitted size ratio of $q^\mathrm{fit}=0.59$ for the smaller polymer (PEG620). Now we consider the assumption that polymers are ideal as in the standard AO model as with $R_g=1.2$ nm, these are \emph{very} small polymers to treat as ideal. \cite{asakura1954} Dijkstra \emph{et al.}\cite{dijkstra1999} compared additive hard-spheres with ideal polymers using thermodynamic perturbation theory, they found that for small \textit{q} and polymer packing fraction $\phi_{p}$, the phase separation is very similar between two models. Here we have $q=0.59$, and under these conditions of larger depletants, the behaviour of spherical colloids plus ideal polymer and spherical colloids plus hard sphere depletant is rather different, at least at the level of the effective interactions between the larger spheres.\cite{royall2007} While we cannot rule out that the polymers may exhibit significant deviations from ideality, given that the phase behavior we find is similar to that of hard spherocylinders and ideal polymer, we note that at the level of our analysis the polymers appear more likely to be behaving in a manner similar to a polymer depletant rather than hard spheres.

\section{Discussion}
\label{sectionDiscussion}

We have seen that the model fluorescent protein--polymer system can, rather surprisingly, be treated in the spirit of a colloid--polymer mixture
where the only additional complexity is an approximate treatment of the anisotropy of the protein dimers. This is notable, and a simple depletion picture of hard spherocylinders with non--absorbing ideal polymers is consistent with our observation. Furthermore, we observe no aggregation for eGFP in the absence of polymer at least to 500 mg/mL, corresponding to a volume fraction of 0.48. At this volume fraction, the protein solution becomes very viscous, consistent with previous work which found glassy behavior reminiscent of colloidal systems in concentrated eye lens $\alpha$-crystallin.\cite{foffi2014} Furthermore, we found that upon dilution aggregated protein solutions re--dissolved, behavior which is compatible with weak, depletion--driven aggregation.

The crystallization behavior in our system re-emphasises that protein crystals can be produced through addition of polymer, as noted previously.\cite{arpino2012} This is significant because the process is apparently immediate without a fine--tuning of the system. We focus on the low volume fraction regime in this work, and we note that crystals only appeared in a limited region in our phase diagram and then only in the system with smaller polymer and added salt, not in the case of the larger polymer or without added salt. At first sight, it may seem surprising that we find above (section \ref{sectionEstimationInteraction}) that the electrostatic interactions are very weak in our system, with or without salt. It is important to highlight that the isoelectric point (pI) of the monomeric unit of the eGFP (obtained from its amino acid sequence) is 5.8, \cite{PepCalc} which is close to the pH 7 used in the experiments. This might explain the small values found for the surface charge.

We now enquire as to why not adding salt suppresses the crystallization. The observation of crystallization only in a very limited region of polymer concentration (i.e. attraction strength) is consistent with previous work with (spherical) colloids and polymer mixtures,\cite{poon2002,royall2012,taylor2012} and has been interpreted in terms of fluctuation--dissipation theorem violation.\cite{klotsa2011} Additionally, it has been observed that acidic proteins are more likely to crystallize when the pH of the solution is 0--2.5 units above their pI. \cite{kirkwood2015} Our experiments fall within such a range. Thus, only a small amount of salt would be required to overcome small electrostatic repulsions under these favourable conditions. What is perhaps more notable is the limited range of protein concentration in which we see crystallization and the failure of the salt--free system to crystallize. It is quite possible that the region of the phase diagram in which crystallization occurs is so small is somehow related to more complex behavior than that which we treat here. For example, Fusco \emph{et al.} showed the importance of contacts in the crystallization behavior of rubredoxin. \cite{fusco2014} We speculate that a decrease in the electrostatic repulsions only needs to occur around or in these regions to promote crystal formation, leading to only small amounts of salt required to yield a crystal, in contrast for example with isotropic systems. Finally, salts can also affect the hydrophobic protein-protein interactions by increasing the surface tension.\cite{durbin1996} These interactions have shown to be relevant in the formation of a crystal phase and protein solubility,\cite{prevost1991,quinn2019} which cannot be discarded in the present study.

Nevertheless, the crystallization that we observe is compatible with the spherocylinder--polymer phase behavior ($L/D-1 = 5$), \cite{savenko2006,bolhuis1997}. It would be most interesting to determine the phase diagram for hard spherocylinders of aspect ratio $L/D-1 = 1.05$ plus polymer, but for now we conclude that our finding of protein crystallization is not inconsistent with some of the literature 
for hard particle -- polymer mixtures.\cite{savenko2006,bolhuis1997,lekkerkerker1992,poon2002,lekkerkerker2011}

The polymer volume fractions at which we find aggregation are rather high, of order unity. It is important to enquire whether one can still apply the concept of polymer--induced depletion under these conditions. Accurate computer simulations in which the polymer chain segments predict that for the polymer fractions that we consider here, only small deviations of ideal Asakura--Oosawa behavior are expected.\cite{louis2002} While we have treated our eGFP as spherocylinders, and this work refers to spherical particles, we are unaware of similar work which pertains to anisotropic particles and thus, in absence of evidence to the contrary, presume that a simple depletion picture remains reasonably accurate at these polymer concentrations.

While we have suggested that it is possible to account for the behavior of our system by treating the eGFP as hard spherocylinders in a solution of ideal polymers, we can be confident that the situation in reality is much more complex. In addition to an enhancement of hydrophobic interactions from salt addition discussed above, due to the amphiphilic nature of PEG, additional hydrophobic \cite{Curtis2006} and chemical \cite{Shkel2015} interactions (via PEG -CH$_2$OCH$_2$- groups) between PEG and proteins might also contribute to this phenomenon. Furthermore, PEG molecules can also enhance aggregation and crystallization via effective repulsion since PEG might preferentially form hydrogen bonds with water compared to the proteins.\cite{durbin1996} Finally, we have determined electrostatic interactions between eGFP dimers to be weak, if we only consider the \emph{net} charge. Of course this is a very significant approximation. Monomeric eGFP has a number of charging groups, e.g. 32 acidic residues and so a more sophisticated approach which takes this into account may prove valuable. Such an approach as that noted above for rubredoxin \cite{fusco2014} would be most interesting to pursue here.

In short, further work is needed to explore throughout the metastable region and then predictions can be validated using the depletion theory. Moreover the properties of those crystals formed at this low protein concentration and by purely depletion interactions, are certainly worth investigating in future research.

\section{Conclusion}
We studied the phase behavior of a model system of fluorescent proteins and polymers (eGFP-PEG) in the ``colloid limit'' where the polymer depletant is smaller than or comparable in size to the protein. A phase behavior of fluid--aggregation was observed for two polymer sizes, i.e. two polymer--protein size ratios). In addition to a small region of the phase diagram of a system with added salt (NaCl) and small polymers where protein crystallization occurred. At high polymer concentration, protein aggregates were large enough to sediment on the timescale of the experiment and form a sediment whose structure is reminiscent of a gel. In the absence of polymer, solutions of eGFP are stable at least to a concentration of 500 mg/ml (volume fraction at 0.48). This suggests that the eGFP dimers interact rather weakly and that approximating them as hard particles may be reasonable.

Based on the shape of eGFP dimers as deduced from small angle x-ray scattering, we treat them as hard spherocylinders with aspect radio $L/D-1 = 1.05$. In the case of the small polymer (PEG 620), the aggregation boundary of polymer fugacity around protein volume fraction of 0.002, was found almost indistinguishable, between $30.0 \pm1.0$ for salt-screened system and $30.9\pm 1.9$ for salt-free system. For the larger polymer (PEG2000) aggregation was found at a polymer fugacity of 1.20. Consistent with DLVO theory for colloids, the effects of electrostatic interactions between the proteins were found to be weak. Intriguingly, in the case of no added salt, and also in the case of no added polymer, we observed no protein crystallization. Due to the uncertainty of the polymer radius of gyration, we interpolated the fugacity for the aggregation phase boundary from existing literature, between $L/D -1= 0$ for spheres--polymer mixtures \cite{loverso2006} and $L/D-1 = 5$ for spherocylinder--polymer mixtures \cite{savenko2006} and fitted a polymer radius of gyration of 1.1 nm for PEG620. Compared with the empirical estimation of 0.83 nm, this somewhat larger size may be related to some non--ideality in the polymers  \cite{lekkerkerker2011} (we note that polymer scaling theory is expected to break down for such small polymers in any case). The smaller difference for larger polymer (PEG2000) is consistent with this.

The behavior we observed is consistent with the depletion picture of hard spherocylinders and ideal polymers. But in reality the system is rather more complex. At our level of analysis and observation, we cannot exclude the possibility that other interactions drive the phenomena that we observe, for example hydration effects, hydrophobic or electrostatic ``patches''. Nevertheless, the fact that in the absence of polymer, the eGFP solution exhibits no aggregation to such high concentrations, at that the aggregates re--dissolve upon dilution gives us some cautious optimism that the behavior we observe may be driven by such simple interactions as the excluded volume effects of polymer--induced depletion.

\begin{acknowledgments}
We would like to thank John Russo and Mike Allen for helpful discussions; Richard Stenner for protein expression and purification; and Ang\'elique Coutable-Pennarun for assistance with zeta potential measurements. This work was financially supported by Bristol Centre for Functional Nanomaterials (BCFN), Chinese Scholarship Council (CSC), and Bayer AG. IRdA was funded by the Philip Leverhulme Prize 2018 awarded by the Leverhulme Trust. IRdA, JLRA and CPR were funded by the Leverhulme Trust grant ``Unifying Protein Design and Assembly of Soft Matter for New Materials''. RC, IRdA and CPR gratefully acknowledge the ERC Grant agreement no. 617266 NANOPRS for financial support and Engineering and Physical Sciences Research Council (EP/H022333/1). The Ganesha X-ray scattering apparatus used for this research was purchased under EPSRC Grant Atoms to Applications (EP/K035746/1).  This work benefitted from the SasView software, originally developed by the DANSE project under NSF award DMR-0520547.
\end{acknowledgments}

\vspace{12pt}
\textbf{Data availability statement}
The data that support the findings of this study are available from the corresponding author upon reasonable request.

\appendix
\section{Geometric Parameters of eGFP Determined with SAXS}

\begin{table}[h!]
\caption{Parameters obtained from SASView \cite{sasview} fitting using a cylinder form factor}
\begin{tabularx}{\columnwidth}{@{}l *6{>{\centering\arraybackslash}X}@{}}
\hline\hline
Protein & Radius ($\AA$) & Error Radius ($\AA$) & Length ($\AA$) & Error Length ($\AA$) & Fitting $\chi^2$\\
\hline
eGFP & 20.5 &0.08 & 82.3 & 0.7 & 1.19\\
\hline\hline
\end{tabularx}
\label{tableSAXSfitting}
\end{table}


\begin{thebibliography}{115}%
\makeatletter
\providecommand \@ifxundefined [1]{%
 \@ifx{#1\undefined}
}%
\providecommand \@ifnum [1]{%
 \ifnum #1\expandafter \@firstoftwo
 \else \expandafter \@secondoftwo
 \fi
}%
\providecommand \@ifx [1]{%
 \ifx #1\expandafter \@firstoftwo
 \else \expandafter \@secondoftwo
 \fi
}%
\providecommand \natexlab [1]{#1}%
\providecommand \enquote  [1]{``#1''}%
\providecommand \bibnamefont  [1]{#1}%
\providecommand \bibfnamefont [1]{#1}%
\providecommand \citenamefont [1]{#1}%
\providecommand \href@noop [0]{\@secondoftwo}%
\providecommand \href [0]{\begingroup \@sanitize@url \@href}%
\providecommand \@href[1]{\@@startlink{#1}\@@href}%
\providecommand \@@href[1]{\endgroup#1\@@endlink}%
\providecommand \@sanitize@url [0]{\catcode `\\12\catcode `\$12\catcode
  `\&12\catcode `\#12\catcode `\^12\catcode `\_12\catcode `\%12\relax}%
\providecommand \@@startlink[1]{}%
\providecommand \@@endlink[0]{}%
\providecommand \url  [0]{\begingroup\@sanitize@url \@url }%
\providecommand \@url [1]{\endgroup\@href {#1}{\urlprefix }}%
\providecommand \urlprefix  [0]{URL }%
\providecommand \Eprint [0]{\href }%
\providecommand \doibase [0]{http://dx.doi.org/}%
\providecommand \selectlanguage [0]{\@gobble}%
\providecommand \bibinfo  [0]{\@secondoftwo}%
\providecommand \bibfield  [0]{\@secondoftwo}%
\providecommand \translation [1]{[#1]}%
\providecommand \BibitemOpen [0]{}%
\providecommand \bibitemStop [0]{}%
\providecommand \bibitemNoStop [0]{.\EOS\space}%
\providecommand \EOS [0]{\spacefactor3000\relax}%
\providecommand \BibitemShut  [1]{\csname bibitem#1\endcsname}%
\let\auto@bib@innerbib\@empty
\bibitem [{\citenamefont {Schneider}, \citenamefont {Rasband},\ and\
  \citenamefont {Eliceiri}(2012)}]{schneider2012}%
  \BibitemOpen
  \bibfield  {author} {\bibinfo {author} {\bibfnamefont {C.~A.}\ \bibnamefont
  {Schneider}}, \bibinfo {author} {\bibfnamefont {W.~S.}\ \bibnamefont
  {Rasband}}, \ and\ \bibinfo {author} {\bibfnamefont {K.~W.}\ \bibnamefont
  {Eliceiri}},\ }\bibfield  {title} {\enquote {\bibinfo {title} {{{NIH Image}}
  to {{ImageJ}}: 25 years of image analysis},}\ }\href {\doibase
  10.1038/nmeth.2089} {\bibfield  {journal} {\bibinfo  {journal} {Nature
  Methods}\ }\textbf {\bibinfo {volume} {9}},\ \bibinfo {pages} {671--675}
  (\bibinfo {year} {2012})}\BibitemShut {NoStop}%
\bibitem [{\citenamefont {Morris}, \citenamefont {Watzky},\ and\ \citenamefont
  {Finke}(2009)}]{morris2009}%
  \BibitemOpen
  \bibfield  {author} {\bibinfo {author} {\bibfnamefont {A.~M.}\ \bibnamefont
  {Morris}}, \bibinfo {author} {\bibfnamefont {M.~A.}\ \bibnamefont {Watzky}},
  \ and\ \bibinfo {author} {\bibfnamefont {R.~G.}\ \bibnamefont {Finke}},\
  }\bibfield  {title} {\enquote {\bibinfo {title} {Protein aggregation
  kinetics, mechanism, and curve-fitting: {{A}} review of the literature},}\
  }\href {\doibase 10.1016/j.bbapap.2008.10.016} {\bibfield  {journal}
  {\bibinfo  {journal} {Biochimica et Biophysica Acta (BBA) - Proteins and
  Proteomics}\ }\textbf {\bibinfo {volume} {1794}},\ \bibinfo {pages}
  {375--397} (\bibinfo {year} {2009})}\BibitemShut {NoStop}%
\bibitem [{\citenamefont {Cohen}\ \emph {et~al.}(2013)\citenamefont {Cohen},
  \citenamefont {Linse}, \citenamefont {Luheshi}, \citenamefont {Hellstrand},
  \citenamefont {White}, \citenamefont {Rajah}, \citenamefont {Otzen},
  \citenamefont {Vendruscolo}, \citenamefont {Dobson},\ and\ \citenamefont
  {Knowles}}]{cohen2013}%
  \BibitemOpen
  \bibfield  {author} {\bibinfo {author} {\bibfnamefont {S.~I.~A.}\
  \bibnamefont {Cohen}}, \bibinfo {author} {\bibfnamefont {S.}~\bibnamefont
  {Linse}}, \bibinfo {author} {\bibfnamefont {L.~M.}\ \bibnamefont {Luheshi}},
  \bibinfo {author} {\bibfnamefont {E.}~\bibnamefont {Hellstrand}}, \bibinfo
  {author} {\bibfnamefont {D.~A.}\ \bibnamefont {White}}, \bibinfo {author}
  {\bibfnamefont {L.}~\bibnamefont {Rajah}}, \bibinfo {author} {\bibfnamefont
  {D.~E.}\ \bibnamefont {Otzen}}, \bibinfo {author} {\bibfnamefont
  {M.}~\bibnamefont {Vendruscolo}}, \bibinfo {author} {\bibfnamefont {C.~M.}\
  \bibnamefont {Dobson}}, \ and\ \bibinfo {author} {\bibfnamefont {T.~P.~J.}\
  \bibnamefont {Knowles}},\ }\bibfield  {title} {\enquote {\bibinfo {title}
  {Proliferation of amyloid- 42 aggregates occurs through a secondary
  nucleation mechanism},}\ }\href {\doibase 10.1073/pnas.1218402110} {\bibfield
   {journal} {\bibinfo  {journal} {Proceedings of the National Academy of
  Sciences}\ }\textbf {\bibinfo {volume} {110}},\ \bibinfo {pages} {9758--9763}
  (\bibinfo {year} {2013})}\BibitemShut {NoStop}%
\bibitem [{\citenamefont {Bai}, \citenamefont {Luo},\ and\ \citenamefont
  {Liu}(2016)}]{bai2016}%
  \BibitemOpen
  \bibfield  {author} {\bibinfo {author} {\bibfnamefont {Y.}~\bibnamefont
  {Bai}}, \bibinfo {author} {\bibfnamefont {Q.}~\bibnamefont {Luo}}, \ and\
  \bibinfo {author} {\bibfnamefont {J.}~\bibnamefont {Liu}},\ }\bibfield
  {title} {\enquote {\bibinfo {title} {Protein self-assembly via supramolecular
  strategies},}\ }\href {\doibase 10.1039/C6CS00004E} {\bibfield  {journal}
  {\bibinfo  {journal} {Chemical Society Reviews}\ }\textbf {\bibinfo {volume}
  {45}},\ \bibinfo {pages} {2756--2767} (\bibinfo {year} {2016})}\BibitemShut
  {NoStop}%
\bibitem [{\citenamefont {Alberti}, \citenamefont {Gladfelter},\ and\
  \citenamefont {Mittag}(2019)}]{alberti2019}%
  \BibitemOpen
  \bibfield  {author} {\bibinfo {author} {\bibfnamefont {S.}~\bibnamefont
  {Alberti}}, \bibinfo {author} {\bibfnamefont {A.}~\bibnamefont {Gladfelter}},
  \ and\ \bibinfo {author} {\bibfnamefont {T.}~\bibnamefont {Mittag}},\
  }\bibfield  {title} {\enquote {\bibinfo {title} {Considerations and
  {{Challenges}} in {{Studying Liquid}}-{{Liquid Phase Separation}} and
  {{Biomolecular Condensates}}},}\ }\href {\doibase 10.1016/j.cell.2018.12.035}
  {\bibfield  {journal} {\bibinfo  {journal} {Cell}\ }\textbf {\bibinfo
  {volume} {176}},\ \bibinfo {pages} {419--434} (\bibinfo {year}
  {2019})}\BibitemShut {NoStop}%
\bibitem [{\citenamefont {Lai}, \citenamefont {King},\ and\ \citenamefont
  {Yeates}(2012)}]{lai2012}%
  \BibitemOpen
  \bibfield  {author} {\bibinfo {author} {\bibfnamefont {Y.-T.}\ \bibnamefont
  {Lai}}, \bibinfo {author} {\bibfnamefont {N.~P.}\ \bibnamefont {King}}, \
  and\ \bibinfo {author} {\bibfnamefont {T.~O.}\ \bibnamefont {Yeates}},\
  }\bibfield  {title} {\enquote {\bibinfo {title} {Principles for designing
  ordered protein assemblies},}\ }\href {\doibase 10.1016/j.tcb.2012.08.004}
  {\bibfield  {journal} {\bibinfo  {journal} {Trends in Cell Biology}\ }\textbf
  {\bibinfo {volume} {22}},\ \bibinfo {pages} {653--661} (\bibinfo {year}
  {2012})}\BibitemShut {NoStop}%
\bibitem [{\citenamefont {Dumetz}\ \emph {et~al.}(2008)\citenamefont {Dumetz},
  \citenamefont {Lewus}, \citenamefont {Lenhoff},\ and\ \citenamefont
  {Kaler}}]{dumetz2008}%
  \BibitemOpen
  \bibfield  {author} {\bibinfo {author} {\bibfnamefont {A.~C.}\ \bibnamefont
  {Dumetz}}, \bibinfo {author} {\bibfnamefont {R.~A.}\ \bibnamefont {Lewus}},
  \bibinfo {author} {\bibfnamefont {A.~M.}\ \bibnamefont {Lenhoff}}, \ and\
  \bibinfo {author} {\bibfnamefont {E.~W.}\ \bibnamefont {Kaler}},\ }\bibfield
  {title} {\enquote {\bibinfo {title} {Effects of {{Ammonium Sulfate}} and
  {{Sodium Chloride Concentration}} on {{PEG}}/{{Protein Liquid}}-{{Liquid
  Phase Separation}}},}\ }\href {\doibase 10.1021/la801180n} {\bibfield
  {journal} {\bibinfo  {journal} {Langmuir}\ }\textbf {\bibinfo {volume}
  {24}},\ \bibinfo {pages} {10345--10351} (\bibinfo {year} {2008})}\BibitemShut
  {NoStop}%
\bibitem [{\citenamefont {Fusco}\ and\ \citenamefont
  {Charbonneau}(2016)}]{fusco2016}%
  \BibitemOpen
  \bibfield  {author} {\bibinfo {author} {\bibfnamefont {D.}~\bibnamefont
  {Fusco}}\ and\ \bibinfo {author} {\bibfnamefont {P.}~\bibnamefont
  {Charbonneau}},\ }\bibfield  {title} {\enquote {\bibinfo {title} {Soft matter
  perspective on protein crystal assembly},}\ }\href {\doibase
  10.1016/j.colsurfb.2015.07.023} {\bibfield  {journal} {\bibinfo  {journal}
  {Colloids and Surfaces B: Biointerfaces}\ }\textbf {\bibinfo {volume}
  {137}},\ \bibinfo {pages} {22--31} (\bibinfo {year} {2016})}\BibitemShut
  {NoStop}%
\bibitem [{\citenamefont {Hui}\ and\ \citenamefont {Edwards}(2003)}]{hui2003}%
  \BibitemOpen
  \bibfield  {author} {\bibinfo {author} {\bibfnamefont {R.}~\bibnamefont
  {Hui}}\ and\ \bibinfo {author} {\bibfnamefont {A.}~\bibnamefont {Edwards}},\
  }\bibfield  {title} {\enquote {\bibinfo {title} {High-throughput protein
  crystallization},}\ }\href@noop {} {\bibfield  {journal} {\bibinfo  {journal}
  {Journal of structural biology}\ }\textbf {\bibinfo {volume} {142}},\
  \bibinfo {pages} {154--161} (\bibinfo {year} {2003})}\BibitemShut {NoStop}%
\bibitem [{\citenamefont {Huang}, \citenamefont {Teng},\ and\ \citenamefont
  {Niu}(1999)}]{huang1999}%
  \BibitemOpen
  \bibfield  {author} {\bibinfo {author} {\bibfnamefont {Q.~q.}\ \bibnamefont
  {Huang}}, \bibinfo {author} {\bibfnamefont {M.~k.}\ \bibnamefont {Teng}}, \
  and\ \bibinfo {author} {\bibfnamefont {L.~w.}\ \bibnamefont {Niu}},\
  }\bibfield  {title} {\enquote {\bibinfo {title} {Protein crystallization with
  a combination of hard and soft precipitants},}\ }\href@noop {} {\bibfield
  {journal} {\bibinfo  {journal} {Acta Crystallographica Section D: Biological
  Crystallography}\ }\textbf {\bibinfo {volume} {55}},\ \bibinfo {pages}
  {1444--1448} (\bibinfo {year} {1999})}\BibitemShut {NoStop}%
\bibitem [{\citenamefont {Bhamidi}, \citenamefont {Varanasi},\ and\
  \citenamefont {Schall}(2005)}]{bhamidi2005}%
  \BibitemOpen
  \bibfield  {author} {\bibinfo {author} {\bibfnamefont {V.}~\bibnamefont
  {Bhamidi}}, \bibinfo {author} {\bibfnamefont {S.}~\bibnamefont {Varanasi}}, \
  and\ \bibinfo {author} {\bibfnamefont {C.~A.}\ \bibnamefont {Schall}},\
  }\bibfield  {title} {\enquote {\bibinfo {title} {Protein {{Crystal
  Nucleation}}: {{Is}} the {{Pair Interaction Potential}} the {{Primary
  Determinant}} of {{Kinetics}}?}}\ }\href {\doibase 10.1021/la050711g}
  {\bibfield  {journal} {\bibinfo  {journal} {Langmuir}\ }\textbf {\bibinfo
  {volume} {21}},\ \bibinfo {pages} {9044--9050} (\bibinfo {year}
  {2005})}\BibitemShut {NoStop}%
\bibitem [{\citenamefont {Glaser}\ and\ \citenamefont
  {Glotzer}(2019)}]{glaser2019}%
  \BibitemOpen
  \bibfield  {author} {\bibinfo {author} {\bibfnamefont {J.}~\bibnamefont
  {Glaser}}\ and\ \bibinfo {author} {\bibfnamefont {S.~C.}\ \bibnamefont
  {Glotzer}},\ }\bibfield  {title} {\enquote {\bibinfo {title} {Looped
  liquid-liquid coexistence in protein crystallization},}\ }\href@noop {}
  {\bibfield  {journal} {\bibinfo  {journal} {arXiv preprint arXiv:1910.06865}\
  } (\bibinfo {year} {2019})}\BibitemShut {NoStop}%
\bibitem [{\citenamefont {Nicolai}\ and\ \citenamefont
  {Durand}(2013)}]{nicolai2013}%
  \BibitemOpen
  \bibfield  {author} {\bibinfo {author} {\bibfnamefont {T.}~\bibnamefont
  {Nicolai}}\ and\ \bibinfo {author} {\bibfnamefont {D.}~\bibnamefont
  {Durand}},\ }\bibfield  {title} {\enquote {\bibinfo {title} {Controlled food
  protein aggregation for new functionality},}\ }\href {\doibase
  10.1016/j.cocis.2013.03.001} {\bibfield  {journal} {\bibinfo  {journal}
  {Current Opinion in Colloid \& Interface Science}\ }\textbf {\bibinfo
  {volume} {18}},\ \bibinfo {pages} {249--256} (\bibinfo {year}
  {2013})}\BibitemShut {NoStop}%
\bibitem [{\citenamefont {Yamazaki}\ \emph {et~al.}(2017)\citenamefont
  {Yamazaki}, \citenamefont {Kimura}, \citenamefont {Vekilov}, \citenamefont
  {Furukawa}, \citenamefont {Shirai}, \citenamefont {Matsumoto}, \citenamefont
  {Van~Driessche},\ and\ \citenamefont {Tsukamoto}}]{yamazaki2017}%
  \BibitemOpen
  \bibfield  {author} {\bibinfo {author} {\bibfnamefont {T.}~\bibnamefont
  {Yamazaki}}, \bibinfo {author} {\bibfnamefont {Y.}~\bibnamefont {Kimura}},
  \bibinfo {author} {\bibfnamefont {P.~G.}\ \bibnamefont {Vekilov}}, \bibinfo
  {author} {\bibfnamefont {E.}~\bibnamefont {Furukawa}}, \bibinfo {author}
  {\bibfnamefont {M.}~\bibnamefont {Shirai}}, \bibinfo {author} {\bibfnamefont
  {H.}~\bibnamefont {Matsumoto}}, \bibinfo {author} {\bibfnamefont {A.~E.}\
  \bibnamefont {Van~Driessche}}, \ and\ \bibinfo {author} {\bibfnamefont
  {K.}~\bibnamefont {Tsukamoto}},\ }\bibfield  {title} {\enquote {\bibinfo
  {title} {Two types of amorphous protein particles facilitate crystal
  nucleation},}\ }\href@noop {} {\bibfield  {journal} {\bibinfo  {journal}
  {Proceedings of the National Academy of Sciences}\ }\textbf {\bibinfo
  {volume} {114}},\ \bibinfo {pages} {2154--2159} (\bibinfo {year}
  {2017})}\BibitemShut {NoStop}%
\bibitem [{\citenamefont {McPherson}(1976)}]{mcpherson1976}%
  \BibitemOpen
  \bibfield  {author} {\bibinfo {author} {\bibfnamefont {A.}~\bibnamefont
  {McPherson}},\ }\bibfield  {title} {\enquote {\bibinfo {title}
  {Crystallization of proteins from polyethylene glycol},}\ }\href {\doibase
  10.1016/S0021-9258(20)81858-4} {\bibfield  {journal} {\bibinfo  {journal}
  {Journal of Biological Chemistry}\ }\textbf {\bibinfo {volume} {251}},\
  \bibinfo {pages} {6300--6303} (\bibinfo {year} {1976})}\BibitemShut {NoStop}%
\bibitem [{\citenamefont {Tanaka}\ and\ \citenamefont
  {Ataka}(2002)}]{tanaka2002}%
  \BibitemOpen
  \bibfield  {author} {\bibinfo {author} {\bibfnamefont {S.}~\bibnamefont
  {Tanaka}}\ and\ \bibinfo {author} {\bibfnamefont {M.}~\bibnamefont {Ataka}},\
  }\bibfield  {title} {\enquote {\bibinfo {title} {Protein crystallization
  induced by polyethylene glycol: {{A}} model study using apoferritin},}\
  }\href {\doibase 10.1063/1.1477456} {\bibfield  {journal} {\bibinfo
  {journal} {The Journal of Chemical Physics}\ }\textbf {\bibinfo {volume}
  {117}},\ \bibinfo {pages} {3504--3510} (\bibinfo {year} {2002})}\BibitemShut
  {NoStop}%
\bibitem [{\citenamefont {Tardieu}\ \emph {et~al.}(2002)\citenamefont
  {Tardieu}, \citenamefont {Bonnet{\'e}}, \citenamefont {Finet},\ and\
  \citenamefont {Vivar{\`e}s}}]{tardieu2002}%
  \BibitemOpen
  \bibfield  {author} {\bibinfo {author} {\bibfnamefont {A.}~\bibnamefont
  {Tardieu}}, \bibinfo {author} {\bibfnamefont {F.}~\bibnamefont
  {Bonnet{\'e}}}, \bibinfo {author} {\bibfnamefont {S.}~\bibnamefont {Finet}},
  \ and\ \bibinfo {author} {\bibfnamefont {D.}~\bibnamefont {Vivar{\`e}s}},\
  }\bibfield  {title} {\enquote {\bibinfo {title} {Understanding salt or
  {{PEG}} induced attractive interactions to crystallize biological
  macromolecules},}\ }\href {\doibase 10.1107/S0907444902014439} {\bibfield
  {journal} {\bibinfo  {journal} {Acta Crystallographica Section D Biological
  Crystallography}\ }\textbf {\bibinfo {volume} {58}},\ \bibinfo {pages}
  {1549--1553} (\bibinfo {year} {2002})}\BibitemShut {NoStop}%
\bibitem [{\citenamefont {Yamanaka}\ \emph {et~al.}(2011)\citenamefont
  {Yamanaka}, \citenamefont {Inaka}, \citenamefont {Furubayashi}, \citenamefont
  {Matsushima}, \citenamefont {Takahashi}, \citenamefont {Tanaka},
  \citenamefont {Sano}, \citenamefont {Sato}, \citenamefont {Kobayashi},\ and\
  \citenamefont {Tanaka}}]{yamanaka2011}%
  \BibitemOpen
  \bibfield  {author} {\bibinfo {author} {\bibfnamefont {M.}~\bibnamefont
  {Yamanaka}}, \bibinfo {author} {\bibfnamefont {K.}~\bibnamefont {Inaka}},
  \bibinfo {author} {\bibfnamefont {N.}~\bibnamefont {Furubayashi}}, \bibinfo
  {author} {\bibfnamefont {M.}~\bibnamefont {Matsushima}}, \bibinfo {author}
  {\bibfnamefont {S.}~\bibnamefont {Takahashi}}, \bibinfo {author}
  {\bibfnamefont {H.}~\bibnamefont {Tanaka}}, \bibinfo {author} {\bibfnamefont
  {S.}~\bibnamefont {Sano}}, \bibinfo {author} {\bibfnamefont {M.}~\bibnamefont
  {Sato}}, \bibinfo {author} {\bibfnamefont {T.}~\bibnamefont {Kobayashi}}, \
  and\ \bibinfo {author} {\bibfnamefont {T.}~\bibnamefont {Tanaka}},\
  }\bibfield  {title} {\enquote {\bibinfo {title} {Optimization of salt
  concentration in peg-based crystallization solutions},}\ }\href@noop {}
  {\bibfield  {journal} {\bibinfo  {journal} {Journal of synchrotron
  radiation}\ }\textbf {\bibinfo {volume} {18}},\ \bibinfo {pages} {84--87}
  (\bibinfo {year} {2011})}\BibitemShut {NoStop}%
\bibitem [{\citenamefont {Velev}, \citenamefont {Kaler},\ and\ \citenamefont
  {Lenhoff}(1998)}]{velev1998}%
  \BibitemOpen
  \bibfield  {author} {\bibinfo {author} {\bibfnamefont {O.}~\bibnamefont
  {Velev}}, \bibinfo {author} {\bibfnamefont {E.}~\bibnamefont {Kaler}}, \ and\
  \bibinfo {author} {\bibfnamefont {A.}~\bibnamefont {Lenhoff}},\ }\bibfield
  {title} {\enquote {\bibinfo {title} {Protein {{Interactions}} in {{Solution
  Characterized}} by {{Light}} and {{Neutron Scattering}}: {{Comparison}} of
  {{Lysozyme}} and {{Chymotrypsinogen}}},}\ }\href {\doibase
  10.1016/S0006-3495(98)77713-6} {\bibfield  {journal} {\bibinfo  {journal}
  {Biophysical Journal}\ }\textbf {\bibinfo {volume} {75}},\ \bibinfo {pages}
  {2682--2697} (\bibinfo {year} {1998})}\BibitemShut {NoStop}%
\bibitem [{\citenamefont {Schmit}\ and\ \citenamefont
  {Dill}(2012)}]{schmit2012}%
  \BibitemOpen
  \bibfield  {author} {\bibinfo {author} {\bibfnamefont {J.~D.}\ \bibnamefont
  {Schmit}}\ and\ \bibinfo {author} {\bibfnamefont {K.}~\bibnamefont {Dill}},\
  }\bibfield  {title} {\enquote {\bibinfo {title} {Growth rates of protein
  crystals},}\ }\href@noop {} {\bibfield  {journal} {\bibinfo  {journal}
  {Journal of the American Chemical Society}\ }\textbf {\bibinfo {volume}
  {134}},\ \bibinfo {pages} {3934--3937} (\bibinfo {year} {2012})}\BibitemShut
  {NoStop}%
\bibitem [{\citenamefont {Astier}\ and\ \citenamefont
  {Veesler}(2008)}]{astier2008}%
  \BibitemOpen
  \bibfield  {author} {\bibinfo {author} {\bibfnamefont {J.-P.}\ \bibnamefont
  {Astier}}\ and\ \bibinfo {author} {\bibfnamefont {S.}~\bibnamefont
  {Veesler}},\ }\bibfield  {title} {\enquote {\bibinfo {title} {Using
  {{Temperature To Crystallize Proteins}}: {{A Mini}}-{{Review}}
  {\textsuperscript{\textdagger}}},}\ }\href {\doibase 10.1021/cg800665b}
  {\bibfield  {journal} {\bibinfo  {journal} {Crystal Growth \& Design}\
  }\textbf {\bibinfo {volume} {8}},\ \bibinfo {pages} {4215--4219} (\bibinfo
  {year} {2008})}\BibitemShut {NoStop}%
\bibitem [{\citenamefont {Grouazel}\ \emph {et~al.}(2006)\citenamefont
  {Grouazel}, \citenamefont {Bonnet{\'e}}, \citenamefont {Astier},
  \citenamefont {Fert{\'e}}, \citenamefont {Perez},\ and\ \citenamefont
  {Veesler}}]{grouazel2006}%
  \BibitemOpen
  \bibfield  {author} {\bibinfo {author} {\bibfnamefont {S.}~\bibnamefont
  {Grouazel}}, \bibinfo {author} {\bibfnamefont {F.}~\bibnamefont
  {Bonnet{\'e}}}, \bibinfo {author} {\bibfnamefont {J.-P.}\ \bibnamefont
  {Astier}}, \bibinfo {author} {\bibfnamefont {N.}~\bibnamefont {Fert{\'e}}},
  \bibinfo {author} {\bibfnamefont {J.}~\bibnamefont {Perez}}, \ and\ \bibinfo
  {author} {\bibfnamefont {S.}~\bibnamefont {Veesler}},\ }\bibfield  {title}
  {\enquote {\bibinfo {title} {Exploring {{Bovine Pancreatic Trypsin Inhibitor
  Phase Transitions}}},}\ }\href {\doibase 10.1021/jp0627123} {\bibfield
  {journal} {\bibinfo  {journal} {The Journal of Physical Chemistry B}\
  }\textbf {\bibinfo {volume} {110}},\ \bibinfo {pages} {19664--19670}
  (\bibinfo {year} {2006})}\BibitemShut {NoStop}%
\bibitem [{\citenamefont {Cie{\'s}lik}\ and\ \citenamefont
  {Derewenda}(2009)}]{cieslik2009}%
  \BibitemOpen
  \bibfield  {author} {\bibinfo {author} {\bibfnamefont {M.}~\bibnamefont
  {Cie{\'s}lik}}\ and\ \bibinfo {author} {\bibfnamefont {Z.~S.}\ \bibnamefont
  {Derewenda}},\ }\bibfield  {title} {\enquote {\bibinfo {title} {The role of
  entropy and polarity in intermolecular contacts in protein crystals},}\
  }\href@noop {} {\bibfield  {journal} {\bibinfo  {journal} {Acta
  Crystallographica Section D: Biological Crystallography}\ }\textbf {\bibinfo
  {volume} {65}},\ \bibinfo {pages} {500--509} (\bibinfo {year}
  {2009})}\BibitemShut {NoStop}%
\bibitem [{\citenamefont {Kim}\ \emph {et~al.}(2015)\citenamefont {Kim},
  \citenamefont {Kim}, \citenamefont {Kim}, \citenamefont {Kim},\ and\
  \citenamefont {Jung}}]{kim2015}%
  \BibitemOpen
  \bibfield  {author} {\bibinfo {author} {\bibfnamefont {Y.~E.}\ \bibnamefont
  {Kim}}, \bibinfo {author} {\bibfnamefont {Y.-n.}\ \bibnamefont {Kim}},
  \bibinfo {author} {\bibfnamefont {J.~A.}\ \bibnamefont {Kim}}, \bibinfo
  {author} {\bibfnamefont {H.~M.}\ \bibnamefont {Kim}}, \ and\ \bibinfo
  {author} {\bibfnamefont {Y.}~\bibnamefont {Jung}},\ }\bibfield  {title}
  {\enquote {\bibinfo {title} {Green fluorescent protein nanopolygons as
  monodisperse supramolecular assemblies of functional proteins with defined
  valency},}\ }\href@noop {} {\bibfield  {journal} {\bibinfo  {journal} {Nature
  communications}\ }\textbf {\bibinfo {volume} {6}},\ \bibinfo {pages} {1--9}
  (\bibinfo {year} {2015})}\BibitemShut {NoStop}%
\bibitem [{\citenamefont {Mandell}\ and\ \citenamefont
  {Kortemme}(2009)}]{mandell2009}%
  \BibitemOpen
  \bibfield  {author} {\bibinfo {author} {\bibfnamefont {D.~J.}\ \bibnamefont
  {Mandell}}\ and\ \bibinfo {author} {\bibfnamefont {T.}~\bibnamefont
  {Kortemme}},\ }\bibfield  {title} {\enquote {\bibinfo {title} {Computer-aided
  design of functional protein interactions},}\ }\href {\doibase
  10.1038/nchembio.251} {\bibfield  {journal} {\bibinfo  {journal} {Nature
  Chemical Biology}\ }\textbf {\bibinfo {volume} {5}},\ \bibinfo {pages}
  {797--807} (\bibinfo {year} {2009})}\BibitemShut {NoStop}%
\bibitem [{\citenamefont {Suzuki}\ \emph {et~al.}(2016)\citenamefont {Suzuki},
  \citenamefont {Cardone}, \citenamefont {Restrepo}, \citenamefont
  {Zavattieri}, \citenamefont {Baker},\ and\ \citenamefont
  {Tezcan}}]{suzuki2016}%
  \BibitemOpen
  \bibfield  {author} {\bibinfo {author} {\bibfnamefont {Y.}~\bibnamefont
  {Suzuki}}, \bibinfo {author} {\bibfnamefont {G.}~\bibnamefont {Cardone}},
  \bibinfo {author} {\bibfnamefont {D.}~\bibnamefont {Restrepo}}, \bibinfo
  {author} {\bibfnamefont {P.~D.}\ \bibnamefont {Zavattieri}}, \bibinfo
  {author} {\bibfnamefont {T.~S.}\ \bibnamefont {Baker}}, \ and\ \bibinfo
  {author} {\bibfnamefont {F.~A.}\ \bibnamefont {Tezcan}},\ }\bibfield  {title}
  {\enquote {\bibinfo {title} {Self-assembly of coherently dynamic, auxetic,
  two-dimensional protein crystals},}\ }\href@noop {} {\bibfield  {journal}
  {\bibinfo  {journal} {Nature}\ }\textbf {\bibinfo {volume} {533}},\ \bibinfo
  {pages} {369--373} (\bibinfo {year} {2016})}\BibitemShut {NoStop}%
\bibitem [{\citenamefont {Zhang}\ \emph {et~al.}(2020)\citenamefont {Zhang},
  \citenamefont {Alberstein}, \citenamefont {De~Yoreo},\ and\ \citenamefont
  {Tezcan}}]{zhang2020}%
  \BibitemOpen
  \bibfield  {author} {\bibinfo {author} {\bibfnamefont {S.}~\bibnamefont
  {Zhang}}, \bibinfo {author} {\bibfnamefont {R.~G.}\ \bibnamefont
  {Alberstein}}, \bibinfo {author} {\bibfnamefont {J.~J.}\ \bibnamefont
  {De~Yoreo}}, \ and\ \bibinfo {author} {\bibfnamefont {F.~A.}\ \bibnamefont
  {Tezcan}},\ }\bibfield  {title} {\enquote {\bibinfo {title} {Assembly of a
  patchy protein into variable 2d lattices via tunable multiscale
  interactions},}\ }\href@noop {} {\bibfield  {journal} {\bibinfo  {journal}
  {Nature communications}\ }\textbf {\bibinfo {volume} {11}},\ \bibinfo {pages}
  {1--12} (\bibinfo {year} {2020})}\BibitemShut {NoStop}%
\bibitem [{\citenamefont {McManus}\ \emph {et~al.}(2016)\citenamefont
  {McManus}, \citenamefont {Charbonneau}, \citenamefont {Zaccarelli},\ and\
  \citenamefont {Asherie}}]{mcmanus2016}%
  \BibitemOpen
  \bibfield  {author} {\bibinfo {author} {\bibfnamefont {J.~J.}\ \bibnamefont
  {McManus}}, \bibinfo {author} {\bibfnamefont {P.}~\bibnamefont
  {Charbonneau}}, \bibinfo {author} {\bibfnamefont {E.}~\bibnamefont
  {Zaccarelli}}, \ and\ \bibinfo {author} {\bibfnamefont {N.}~\bibnamefont
  {Asherie}},\ }\bibfield  {title} {\enquote {\bibinfo {title} {The physics of
  protein self-assembly},}\ }\href@noop {} {\bibfield  {journal} {\bibinfo
  {journal} {Current opinion in colloid \& interface science}\ }\textbf
  {\bibinfo {volume} {22}},\ \bibinfo {pages} {73--79} (\bibinfo {year}
  {2016})}\BibitemShut {NoStop}%
\bibitem [{\citenamefont {Stradner}\ and\ \citenamefont
  {Schurtenberger}(2020)}]{stradner2020}%
  \BibitemOpen
  \bibfield  {author} {\bibinfo {author} {\bibfnamefont {A.}~\bibnamefont
  {Stradner}}\ and\ \bibinfo {author} {\bibfnamefont {P.}~\bibnamefont
  {Schurtenberger}},\ }\bibfield  {title} {\enquote {\bibinfo {title}
  {Potential and limits of a colloid approach to protein solutions},}\
  }\href@noop {} {\bibfield  {journal} {\bibinfo  {journal} {Soft Matter}\
  }\textbf {\bibinfo {volume} {16}},\ \bibinfo {pages} {307--323} (\bibinfo
  {year} {2020})}\BibitemShut {NoStop}%
\bibitem [{\citenamefont {Hamley}\ and\ \citenamefont
  {Castelletto}(2007)}]{hamley2007}%
  \BibitemOpen
  \bibfield  {author} {\bibinfo {author} {\bibfnamefont {I.~W.}\ \bibnamefont
  {Hamley}}\ and\ \bibinfo {author} {\bibfnamefont {V.}~\bibnamefont
  {Castelletto}},\ }\bibfield  {title} {\enquote {\bibinfo {title} {Biological
  {{Soft Materials}}},}\ }\href {\doibase 10.1002/anie.200603922} {\bibfield
  {journal} {\bibinfo  {journal} {Angewandte Chemie International Edition}\
  }\textbf {\bibinfo {volume} {46}},\ \bibinfo {pages} {4442--4455} (\bibinfo
  {year} {2007})}\BibitemShut {NoStop}%
\bibitem [{\citenamefont {Dorsaz}\ \emph {et~al.}(2012)\citenamefont {Dorsaz},
  \citenamefont {Filion}, \citenamefont {Smallenburg},\ and\ \citenamefont
  {Frenkel}}]{dorsaz2012}%
  \BibitemOpen
  \bibfield  {author} {\bibinfo {author} {\bibfnamefont {N.}~\bibnamefont
  {Dorsaz}}, \bibinfo {author} {\bibfnamefont {L.}~\bibnamefont {Filion}},
  \bibinfo {author} {\bibfnamefont {F.}~\bibnamefont {Smallenburg}}, \ and\
  \bibinfo {author} {\bibfnamefont {D.}~\bibnamefont {Frenkel}},\ }\bibfield
  {title} {\enquote {\bibinfo {title} {Spiers memorial lecture: Effect of
  interaction specificity on the phase behaviour of patchy particles},}\
  }\href@noop {} {\bibfield  {journal} {\bibinfo  {journal} {Faraday
  discussions}\ }\textbf {\bibinfo {volume} {159}},\ \bibinfo {pages} {9--21}
  (\bibinfo {year} {2012})}\BibitemShut {NoStop}%
\bibitem [{\citenamefont {Zhang}\ \emph {et~al.}(2012)\citenamefont {Zhang},
  \citenamefont {Roth}, \citenamefont {Wolf}, \citenamefont {Roosen-Runge},
  \citenamefont {Skoda}, \citenamefont {Jacobs}, \citenamefont {Stzucki},\ and\
  \citenamefont {Schreiber}}]{zhang2012}%
  \BibitemOpen
  \bibfield  {author} {\bibinfo {author} {\bibfnamefont {F.}~\bibnamefont
  {Zhang}}, \bibinfo {author} {\bibfnamefont {R.}~\bibnamefont {Roth}},
  \bibinfo {author} {\bibfnamefont {M.}~\bibnamefont {Wolf}}, \bibinfo {author}
  {\bibfnamefont {F.}~\bibnamefont {Roosen-Runge}}, \bibinfo {author}
  {\bibfnamefont {M.~W.}\ \bibnamefont {Skoda}}, \bibinfo {author}
  {\bibfnamefont {R.~M.}\ \bibnamefont {Jacobs}}, \bibinfo {author}
  {\bibfnamefont {M.}~\bibnamefont {Stzucki}}, \ and\ \bibinfo {author}
  {\bibfnamefont {F.}~\bibnamefont {Schreiber}},\ }\bibfield  {title} {\enquote
  {\bibinfo {title} {Charge-controlled metastable liquid--liquid phase
  separation in protein solutions as a universal pathway towards
  crystallization},}\ }\href@noop {} {\bibfield  {journal} {\bibinfo  {journal}
  {Soft Matter}\ }\textbf {\bibinfo {volume} {8}},\ \bibinfo {pages}
  {1313--1316} (\bibinfo {year} {2012})}\BibitemShut {NoStop}%
\bibitem [{\citenamefont {Zhang}(2017)}]{zhang2017}%
  \BibitemOpen
  \bibfield  {author} {\bibinfo {author} {\bibfnamefont {F.}~\bibnamefont
  {Zhang}},\ }\bibfield  {title} {\enquote {\bibinfo {title} {Nonclassical
  nucleation pathways in protein crystallization},}\ }\href@noop {} {\bibfield
  {journal} {\bibinfo  {journal} {Journal of Physics: Condensed Matter}\
  }\textbf {\bibinfo {volume} {29}},\ \bibinfo {pages} {443002} (\bibinfo
  {year} {2017})}\BibitemShut {NoStop}%
\bibitem [{\citenamefont {Piazza}(2004)}]{piazza2004}%
  \BibitemOpen
  \bibfield  {author} {\bibinfo {author} {\bibfnamefont {R.}~\bibnamefont
  {Piazza}},\ }\bibfield  {title} {\enquote {\bibinfo {title} {Protein
  interactions and association: an open challenge for colloid science},}\
  }\href@noop {} {\bibfield  {journal} {\bibinfo  {journal} {Current opinion in
  colloid \& interface science}\ }\textbf {\bibinfo {volume} {8}},\ \bibinfo
  {pages} {515--522} (\bibinfo {year} {2004})}\BibitemShut {NoStop}%
\bibitem [{\citenamefont {Tessier}\ \emph {et~al.}(2002)\citenamefont
  {Tessier}, \citenamefont {Johnson}, \citenamefont {Pazhianur}, \citenamefont
  {Berger}, \citenamefont {Prentice}, \citenamefont {Bahnson}, \citenamefont
  {Sandler},\ and\ \citenamefont {Lenhoff}}]{tessier2002}%
  \BibitemOpen
  \bibfield  {author} {\bibinfo {author} {\bibfnamefont {P.~M.}\ \bibnamefont
  {Tessier}}, \bibinfo {author} {\bibfnamefont {H.~R.}\ \bibnamefont
  {Johnson}}, \bibinfo {author} {\bibfnamefont {R.}~\bibnamefont {Pazhianur}},
  \bibinfo {author} {\bibfnamefont {B.~W.}\ \bibnamefont {Berger}}, \bibinfo
  {author} {\bibfnamefont {J.~L.}\ \bibnamefont {Prentice}}, \bibinfo {author}
  {\bibfnamefont {B.~J.}\ \bibnamefont {Bahnson}}, \bibinfo {author}
  {\bibfnamefont {S.~I.}\ \bibnamefont {Sandler}}, \ and\ \bibinfo {author}
  {\bibfnamefont {A.~M.}\ \bibnamefont {Lenhoff}},\ }\bibfield  {title}
  {\enquote {\bibinfo {title} {Predictive crystallization of ribonuclease {{A}}
  via rapid screening of osmotic second virial coefficients},}\ }\href
  {\doibase 10.1002/prot.10249} {\bibfield  {journal} {\bibinfo  {journal}
  {Proteins: Structure, Function, and Bioinformatics}\ }\textbf {\bibinfo
  {volume} {50}},\ \bibinfo {pages} {303--311} (\bibinfo {year}
  {2002})}\BibitemShut {NoStop}%
\bibitem [{\citenamefont {George}\ and\ \citenamefont
  {Wilson}(1994{\natexlab{a}})}]{george1994}%
  \BibitemOpen
  \bibfield  {author} {\bibinfo {author} {\bibfnamefont {A.}~\bibnamefont
  {George}}\ and\ \bibinfo {author} {\bibfnamefont {W.~W.}\ \bibnamefont
  {Wilson}},\ }\bibfield  {title} {\enquote {\bibinfo {title} {Predicting
  protein crystallization from a dilute solution property},}\ }\href@noop {}
  {\bibfield  {journal} {\bibinfo  {journal} {Acta Crystallographica Section D:
  Biological Crystallography}\ }\textbf {\bibinfo {volume} {50}},\ \bibinfo
  {pages} {361--365} (\bibinfo {year} {1994}{\natexlab{a}})}\BibitemShut
  {NoStop}%
\bibitem [{\citenamefont {Elcock}\ and\ \citenamefont
  {McCammon}(2001)}]{elcock2001}%
  \BibitemOpen
  \bibfield  {author} {\bibinfo {author} {\bibfnamefont {A.~H.}\ \bibnamefont
  {Elcock}}\ and\ \bibinfo {author} {\bibfnamefont {J.~A.}\ \bibnamefont
  {McCammon}},\ }\bibfield  {title} {\enquote {\bibinfo {title} {Calculation of
  weak protein-protein interactions: the ph dependence of the second virial
  coefficient},}\ }\href@noop {} {\bibfield  {journal} {\bibinfo  {journal}
  {Biophysical journal}\ }\textbf {\bibinfo {volume} {80}},\ \bibinfo {pages}
  {613--625} (\bibinfo {year} {2001})}\BibitemShut {NoStop}%
\bibitem [{\citenamefont {Wentzel}\ and\ \citenamefont
  {Gunton}(2008)}]{wentzel2008}%
  \BibitemOpen
  \bibfield  {author} {\bibinfo {author} {\bibfnamefont {N.}~\bibnamefont
  {Wentzel}}\ and\ \bibinfo {author} {\bibfnamefont {J.~D.}\ \bibnamefont
  {Gunton}},\ }\bibfield  {title} {\enquote {\bibinfo {title} {Effect of
  solvent on the phase diagram of a simple anisotropic model of globular
  proteins},}\ }\href@noop {} {\bibfield  {journal} {\bibinfo  {journal} {The
  Journal of Physical Chemistry B}\ }\textbf {\bibinfo {volume} {112}},\
  \bibinfo {pages} {7803--7809} (\bibinfo {year} {2008})}\BibitemShut {NoStop}%
\bibitem [{\citenamefont {Noro}\ and\ \citenamefont
  {Frenkel}(2000)}]{noro2000}%
  \BibitemOpen
  \bibfield  {author} {\bibinfo {author} {\bibfnamefont {M.~G.}\ \bibnamefont
  {Noro}}\ and\ \bibinfo {author} {\bibfnamefont {D.}~\bibnamefont {Frenkel}},\
  }\bibfield  {title} {\enquote {\bibinfo {title} {Extended
  corresponding-states behavior for particles with variable range
  attractions},}\ }\href {\doibase 10.1063/1.1288684} {\bibfield  {journal}
  {\bibinfo  {journal} {The Journal of Chemical Physics}\ }\textbf {\bibinfo
  {volume} {113}},\ \bibinfo {pages} {2941--2944} (\bibinfo {year}
  {2000})}\BibitemShut {NoStop}%
\bibitem [{\citenamefont {Lu}\ \emph {et~al.}(2008)\citenamefont {Lu},
  \citenamefont {Zaccarelli}, \citenamefont {Ciulla}, \citenamefont
  {Schofield}, \citenamefont {Sciortino},\ and\ \citenamefont
  {Weitz}}]{lu2008}%
  \BibitemOpen
  \bibfield  {author} {\bibinfo {author} {\bibfnamefont {P.~J.}\ \bibnamefont
  {Lu}}, \bibinfo {author} {\bibfnamefont {E.}~\bibnamefont {Zaccarelli}},
  \bibinfo {author} {\bibfnamefont {F.}~\bibnamefont {Ciulla}}, \bibinfo
  {author} {\bibfnamefont {A.~B.}\ \bibnamefont {Schofield}}, \bibinfo {author}
  {\bibfnamefont {F.}~\bibnamefont {Sciortino}}, \ and\ \bibinfo {author}
  {\bibfnamefont {D.~A.}\ \bibnamefont {Weitz}},\ }\bibfield  {title} {\enquote
  {\bibinfo {title} {Gelation of particles with short-range attraction},}\
  }\href {\doibase 10.1038/nature06931} {\bibfield  {journal} {\bibinfo
  {journal} {Nature}\ }\textbf {\bibinfo {volume} {453}},\ \bibinfo {pages}
  {499--503} (\bibinfo {year} {2008})}\BibitemShut {NoStop}%
\bibitem [{\citenamefont {Royall}(2018)}]{royall2018}%
  \BibitemOpen
  \bibfield  {author} {\bibinfo {author} {\bibfnamefont {C.~P.}\ \bibnamefont
  {Royall}},\ }\bibfield  {title} {\enquote {\bibinfo {title} {Hunting mermaids
  in real space: Known knowns, known unknowns and unknown unknowns},}\ }\href
  {\doibase 10.1039/C8SM00400E} {\bibfield  {journal} {\bibinfo  {journal}
  {Soft Matter}\ }\textbf {\bibinfo {volume} {14}},\ \bibinfo {pages}
  {4020--4028} (\bibinfo {year} {2018})}\BibitemShut {NoStop}%
\bibitem [{\citenamefont {Zhang}\ \emph {et~al.}(2008)\citenamefont {Zhang},
  \citenamefont {Skoda}, \citenamefont {Jacobs}, \citenamefont {Zorn},
  \citenamefont {Martin}, \citenamefont {Martin}, \citenamefont {Clark},
  \citenamefont {Weggler}, \citenamefont {Hildebrandt}, \citenamefont
  {Kohlbacher} \emph {et~al.}}]{zhang2008}%
  \BibitemOpen
  \bibfield  {author} {\bibinfo {author} {\bibfnamefont {F.}~\bibnamefont
  {Zhang}}, \bibinfo {author} {\bibfnamefont {M.}~\bibnamefont {Skoda}},
  \bibinfo {author} {\bibfnamefont {R.}~\bibnamefont {Jacobs}}, \bibinfo
  {author} {\bibfnamefont {S.}~\bibnamefont {Zorn}}, \bibinfo {author}
  {\bibfnamefont {R.~A.}\ \bibnamefont {Martin}}, \bibinfo {author}
  {\bibfnamefont {C.}~\bibnamefont {Martin}}, \bibinfo {author} {\bibfnamefont
  {G.}~\bibnamefont {Clark}}, \bibinfo {author} {\bibfnamefont
  {S.}~\bibnamefont {Weggler}}, \bibinfo {author} {\bibfnamefont
  {A.}~\bibnamefont {Hildebrandt}}, \bibinfo {author} {\bibfnamefont
  {O.}~\bibnamefont {Kohlbacher}},  \emph {et~al.},\ }\bibfield  {title}
  {\enquote {\bibinfo {title} {Reentrant condensation of proteins in solution
  induced by multivalent counterions},}\ }\href@noop {} {\bibfield  {journal}
  {\bibinfo  {journal} {Physical review letters}\ }\textbf {\bibinfo {volume}
  {101}},\ \bibinfo {pages} {148101} (\bibinfo {year} {2008})}\BibitemShut
  {NoStop}%
\bibitem [{\citenamefont {Foffi}\ \emph {et~al.}(2014)\citenamefont {Foffi},
  \citenamefont {Savin}, \citenamefont {Bucciarelli}, \citenamefont {Dorsaz},
  \citenamefont {Thurston}, \citenamefont {Stradner},\ and\ \citenamefont
  {Schurtenberger}}]{foffi2014}%
  \BibitemOpen
  \bibfield  {author} {\bibinfo {author} {\bibfnamefont {G.}~\bibnamefont
  {Foffi}}, \bibinfo {author} {\bibfnamefont {G.}~\bibnamefont {Savin}},
  \bibinfo {author} {\bibfnamefont {S.}~\bibnamefont {Bucciarelli}}, \bibinfo
  {author} {\bibfnamefont {N.}~\bibnamefont {Dorsaz}}, \bibinfo {author}
  {\bibfnamefont {G.~M.}\ \bibnamefont {Thurston}}, \bibinfo {author}
  {\bibfnamefont {A.}~\bibnamefont {Stradner}}, \ and\ \bibinfo {author}
  {\bibfnamefont {P.}~\bibnamefont {Schurtenberger}},\ }\bibfield  {title}
  {\enquote {\bibinfo {title} {Hard sphere-like glass transition in eye lens
  $\alpha$-crystallin solutions},}\ }\href@noop {} {\bibfield  {journal}
  {\bibinfo  {journal} {Proc. Nat. Acad. Sci.}\ }\textbf {\bibinfo {volume}
  {111}},\ \bibinfo {pages} {16748--16753} (\bibinfo {year}
  {2014})}\BibitemShut {NoStop}%
\bibitem [{\citenamefont {Kim}, \citenamefont {Dumont},\ and\ \citenamefont
  {Gruebele}(2008)}]{kim2008}%
  \BibitemOpen
  \bibfield  {author} {\bibinfo {author} {\bibfnamefont {S.~J.}\ \bibnamefont
  {Kim}}, \bibinfo {author} {\bibfnamefont {C.}~\bibnamefont {Dumont}}, \ and\
  \bibinfo {author} {\bibfnamefont {M.}~\bibnamefont {Gruebele}},\ }\bibfield
  {title} {\enquote {\bibinfo {title} {Simulation-based fitting of
  protein-protein interaction potentials to saxs experiments},}\ }\href@noop {}
  {\bibfield  {journal} {\bibinfo  {journal} {Biophysical journal}\ }\textbf
  {\bibinfo {volume} {94}},\ \bibinfo {pages} {4924--4931} (\bibinfo {year}
  {2008})}\BibitemShut {NoStop}%
\bibitem [{\citenamefont {Huang}, \citenamefont {Yao},\ and\ \citenamefont
  {Olsen}(2019)}]{huang2019}%
  \BibitemOpen
  \bibfield  {author} {\bibinfo {author} {\bibfnamefont {A.}~\bibnamefont
  {Huang}}, \bibinfo {author} {\bibfnamefont {H.}~\bibnamefont {Yao}}, \ and\
  \bibinfo {author} {\bibfnamefont {B.~D.}\ \bibnamefont {Olsen}},\ }\bibfield
  {title} {\enquote {\bibinfo {title} {{{SANS}} partial structure factor
  analysis for determining protein\textendash polymer interactions in
  semidilute solution},}\ }\href {\doibase 10.1039/C9SM00766K} {\bibfield
  {journal} {\bibinfo  {journal} {Soft Matter}\ }\textbf {\bibinfo {volume}
  {15}},\ \bibinfo {pages} {7350--7359} (\bibinfo {year} {2019})}\BibitemShut
  {NoStop}%
\bibitem [{\citenamefont {Ten~Wolde}\ and\ \citenamefont
  {Frenkel}(1997)}]{tenwolde1997}%
  \BibitemOpen
  \bibfield  {author} {\bibinfo {author} {\bibfnamefont {P.~R.}\ \bibnamefont
  {Ten~Wolde}}\ and\ \bibinfo {author} {\bibfnamefont {D.}~\bibnamefont
  {Frenkel}},\ }\bibfield  {title} {\enquote {\bibinfo {title} {Enhancement of
  protein crystal nucleation by critical density fluctuations},}\ }\href
  {\doibase 10.1126/science.277.5334.1975} {\bibfield  {journal} {\bibinfo
  {journal} {Science}\ }\textbf {\bibinfo {volume} {277}},\ \bibinfo {pages}
  {1975--1978} (\bibinfo {year} {1997})}\BibitemShut {NoStop}%
\bibitem [{\citenamefont {Pellicane}, \citenamefont {Costa},\ and\
  \citenamefont {Caccamo}(2004)}]{pellicane2004}%
  \BibitemOpen
  \bibfield  {author} {\bibinfo {author} {\bibfnamefont {G.}~\bibnamefont
  {Pellicane}}, \bibinfo {author} {\bibfnamefont {D.}~\bibnamefont {Costa}}, \
  and\ \bibinfo {author} {\bibfnamefont {C.}~\bibnamefont {Caccamo}},\
  }\bibfield  {title} {\enquote {\bibinfo {title} {Theory and simulation of
  short-range models of globular protein solutions},}\ }\href {\doibase
  10.1088/0953-8984/16/42/010} {\bibfield  {journal} {\bibinfo  {journal}
  {Journal of Physics: Condensed Matter}\ }\textbf {\bibinfo {volume} {16}},\
  \bibinfo {pages} {S4923--S4936} (\bibinfo {year} {2004})}\BibitemShut
  {NoStop}%
\bibitem [{\citenamefont {Savage}\ and\ \citenamefont
  {Dinsmore}(2009)}]{savage2009}%
  \BibitemOpen
  \bibfield  {author} {\bibinfo {author} {\bibfnamefont {J.~R.}\ \bibnamefont
  {Savage}}\ and\ \bibinfo {author} {\bibfnamefont {A.~D.}\ \bibnamefont
  {Dinsmore}},\ }\bibfield  {title} {\enquote {\bibinfo {title} {Experimental
  {{Evidence}} for {{Two}}-{{Step Nucleation}} in {{Colloidal
  Crystallization}}},}\ }\href {\doibase 10.1103/PhysRevLett.102.198302}
  {\bibfield  {journal} {\bibinfo  {journal} {Physical Review Letters}\
  }\textbf {\bibinfo {volume} {102}},\ \bibinfo {pages} {198302} (\bibinfo
  {year} {2009})}\BibitemShut {NoStop}%
\bibitem [{\citenamefont {Taylor}, \citenamefont {Evans},\ and\ \citenamefont
  {Royall}(2012)}]{taylor2012}%
  \BibitemOpen
  \bibfield  {author} {\bibinfo {author} {\bibfnamefont {S.~L.}\ \bibnamefont
  {Taylor}}, \bibinfo {author} {\bibfnamefont {R.}~\bibnamefont {Evans}}, \
  and\ \bibinfo {author} {\bibfnamefont {C.~P.}\ \bibnamefont {Royall}},\
  }\bibfield  {title} {\enquote {\bibinfo {title} {Temperature as an external
  field for colloid--polymer mixtures:‘quenching’by heating and
  ‘melting’by cooling},}\ }\href@noop {} {\bibfield  {journal} {\bibinfo
  {journal} {Journal of Physics: Condensed Matter}\ }\textbf {\bibinfo {volume}
  {24}},\ \bibinfo {pages} {464128} (\bibinfo {year} {2012})}\BibitemShut
  {NoStop}%
\bibitem [{\citenamefont {Cardinaux}\ \emph {et~al.}(2007)\citenamefont
  {Cardinaux}, \citenamefont {Gibaud}, \citenamefont {Stradner},\ and\
  \citenamefont {Schurtenberger}}]{cardinaux2007}%
  \BibitemOpen
  \bibfield  {author} {\bibinfo {author} {\bibfnamefont {F.}~\bibnamefont
  {Cardinaux}}, \bibinfo {author} {\bibfnamefont {T.}~\bibnamefont {Gibaud}},
  \bibinfo {author} {\bibfnamefont {A.}~\bibnamefont {Stradner}}, \ and\
  \bibinfo {author} {\bibfnamefont {P.}~\bibnamefont {Schurtenberger}},\
  }\bibfield  {title} {\enquote {\bibinfo {title} {Interplay between {{Spinodal
  Decomposition}} and {{Glass Formation}} in {{Proteins Exhibiting
  Short}}-{{Range Attractions}}},}\ }\href {\doibase
  10.1103/PhysRevLett.99.118301} {\bibfield  {journal} {\bibinfo  {journal}
  {Physical Review Letters}\ }\textbf {\bibinfo {volume} {99}},\ \bibinfo
  {pages} {118301} (\bibinfo {year} {2007})}\BibitemShut {NoStop}%
\bibitem [{\citenamefont {Kulkarni}, \citenamefont {Dixit},\ and\ \citenamefont
  {Zukoski}(2003)}]{kulkarni2003}%
  \BibitemOpen
  \bibfield  {author} {\bibinfo {author} {\bibfnamefont {A.~M.}\ \bibnamefont
  {Kulkarni}}, \bibinfo {author} {\bibfnamefont {N.~M.}\ \bibnamefont {Dixit}},
  \ and\ \bibinfo {author} {\bibfnamefont {C.~F.}\ \bibnamefont {Zukoski}},\
  }\bibfield  {title} {\enquote {\bibinfo {title} {Ergodic and non-ergodic
  phase transitions in globular protein suspensions},}\ }\href@noop {}
  {\bibfield  {journal} {\bibinfo  {journal} {Faraday discussions}\ }\textbf
  {\bibinfo {volume} {123}},\ \bibinfo {pages} {37--50} (\bibinfo {year}
  {2003})}\BibitemShut {NoStop}%
\bibitem [{\citenamefont {Broide}\ \emph {et~al.}(1991)\citenamefont {Broide},
  \citenamefont {Berland}, \citenamefont {Pande}, \citenamefont {Ogun},\ and\
  \citenamefont {Benedek}}]{broide1991}%
  \BibitemOpen
  \bibfield  {author} {\bibinfo {author} {\bibfnamefont {M.~L.}\ \bibnamefont
  {Broide}}, \bibinfo {author} {\bibfnamefont {C.~R.}\ \bibnamefont {Berland}},
  \bibinfo {author} {\bibfnamefont {J.}~\bibnamefont {Pande}}, \bibinfo
  {author} {\bibfnamefont {O.~O.}\ \bibnamefont {Ogun}}, \ and\ \bibinfo
  {author} {\bibfnamefont {G.~B.}\ \bibnamefont {Benedek}},\ }\bibfield
  {title} {\enquote {\bibinfo {title} {Binary-liquid phase separation of lens
  protein solutions.}}\ }\href@noop {} {\bibfield  {journal} {\bibinfo
  {journal} {Proceedings of the National Academy of Sciences}\ }\textbf
  {\bibinfo {volume} {88}},\ \bibinfo {pages} {5660--5664} (\bibinfo {year}
  {1991})}\BibitemShut {NoStop}%
\bibitem [{\citenamefont {Whitelam}(2010)}]{whitelam2010}%
  \BibitemOpen
  \bibfield  {author} {\bibinfo {author} {\bibfnamefont {S.}~\bibnamefont
  {Whitelam}},\ }\bibfield  {title} {\enquote {\bibinfo {title} {Control of
  pathways and yields of protein crystallization through the interplay of
  nonspecific and specific attractions},}\ }\href@noop {} {\bibfield  {journal}
  {\bibinfo  {journal} {Physical review letters}\ }\textbf {\bibinfo {volume}
  {105}},\ \bibinfo {pages} {088102} (\bibinfo {year} {2010})}\BibitemShut
  {NoStop}%
\bibitem [{\citenamefont {Groenewold}\ and\ \citenamefont
  {Kegel}(2001)}]{groenewold2001}%
  \BibitemOpen
  \bibfield  {author} {\bibinfo {author} {\bibfnamefont {J.}~\bibnamefont
  {Groenewold}}\ and\ \bibinfo {author} {\bibfnamefont {W.~K.}\ \bibnamefont
  {Kegel}},\ }\bibfield  {title} {\enquote {\bibinfo {title} {Anomalously
  {{Large Equilibrium Clusters}} of {{Colloids}}
  {\textsuperscript{\textdagger}}},}\ }\href {\doibase 10.1021/jp011646w}
  {\bibfield  {journal} {\bibinfo  {journal} {The Journal of Physical Chemistry
  B}\ }\textbf {\bibinfo {volume} {105}},\ \bibinfo {pages} {11702--11709}
  (\bibinfo {year} {2001})}\BibitemShut {NoStop}%
\bibitem [{\citenamefont {Stradner}\ \emph {et~al.}(2004)\citenamefont
  {Stradner}, \citenamefont {Sedgwick}, \citenamefont {Cardinaux},
  \citenamefont {Poon}, \citenamefont {Egelhaaf},\ and\ \citenamefont
  {Schurtenberger}}]{stradner2004}%
  \BibitemOpen
  \bibfield  {author} {\bibinfo {author} {\bibfnamefont {A.}~\bibnamefont
  {Stradner}}, \bibinfo {author} {\bibfnamefont {H.}~\bibnamefont {Sedgwick}},
  \bibinfo {author} {\bibfnamefont {F.}~\bibnamefont {Cardinaux}}, \bibinfo
  {author} {\bibfnamefont {W.~C.~K.}\ \bibnamefont {Poon}}, \bibinfo {author}
  {\bibfnamefont {S.~U.}\ \bibnamefont {Egelhaaf}}, \ and\ \bibinfo {author}
  {\bibfnamefont {P.}~\bibnamefont {Schurtenberger}},\ }\bibfield  {title}
  {\enquote {\bibinfo {title} {Equilibrium cluster formation in concentrated
  protein solutions and colloids},}\ }\href {\doibase 10.1038/nature03109}
  {\bibfield  {journal} {\bibinfo  {journal} {Nature}\ }\textbf {\bibinfo
  {volume} {432}},\ \bibinfo {pages} {492--495} (\bibinfo {year}
  {2004})}\BibitemShut {NoStop}%
\bibitem [{\citenamefont {Sedgwick}, \citenamefont {Egelhaaf},\ and\
  \citenamefont {Poon}(2004)}]{sedgwick2004}%
  \BibitemOpen
  \bibfield  {author} {\bibinfo {author} {\bibfnamefont {H.}~\bibnamefont
  {Sedgwick}}, \bibinfo {author} {\bibfnamefont {S.~U.}\ \bibnamefont
  {Egelhaaf}}, \ and\ \bibinfo {author} {\bibfnamefont {W.~C.~K.}\ \bibnamefont
  {Poon}},\ }\bibfield  {title} {\enquote {\bibinfo {title} {Clusters and gels
  in systems of sticky particles},}\ }\href {\doibase
  10.1088/0953-8984/16/42/009} {\bibfield  {journal} {\bibinfo  {journal}
  {Journal of Physics: Condensed Matter}\ }\textbf {\bibinfo {volume} {16}},\
  \bibinfo {pages} {S4913--S4922} (\bibinfo {year} {2004})}\BibitemShut
  {NoStop}%
\bibitem [{\citenamefont {Campbell}\ \emph {et~al.}(2005)\citenamefont
  {Campbell}, \citenamefont {Anderson}, \citenamefont {{van Duijneveldt}},\
  and\ \citenamefont {Bartlett}}]{campbell2005}%
  \BibitemOpen
  \bibfield  {author} {\bibinfo {author} {\bibfnamefont {A.~I.}\ \bibnamefont
  {Campbell}}, \bibinfo {author} {\bibfnamefont {V.~J.}\ \bibnamefont
  {Anderson}}, \bibinfo {author} {\bibfnamefont {J.~S.}\ \bibnamefont {{van
  Duijneveldt}}}, \ and\ \bibinfo {author} {\bibfnamefont {P.}~\bibnamefont
  {Bartlett}},\ }\bibfield  {title} {\enquote {\bibinfo {title} {Dynamical
  {{Arrest}} in {{Attractive Colloids}}: {{The Effect}} of {{Long}}-{{Range
  Repulsion}}},}\ }\href {\doibase 10.1103/PhysRevLett.94.208301} {\bibfield
  {journal} {\bibinfo  {journal} {Physical Review Letters}\ }\textbf {\bibinfo
  {volume} {94}},\ \bibinfo {pages} {208301} (\bibinfo {year}
  {2005})}\BibitemShut {NoStop}%
\bibitem [{\citenamefont {Sciortino}, \citenamefont {Tartaglia},\ and\
  \citenamefont {Zaccarelli}(2005)}]{sciortino2005}%
  \BibitemOpen
  \bibfield  {author} {\bibinfo {author} {\bibfnamefont {F.}~\bibnamefont
  {Sciortino}}, \bibinfo {author} {\bibfnamefont {P.}~\bibnamefont
  {Tartaglia}}, \ and\ \bibinfo {author} {\bibfnamefont {E.}~\bibnamefont
  {Zaccarelli}},\ }\bibfield  {title} {\enquote {\bibinfo {title}
  {One-{{Dimensional Cluster Growth}} and {{Branching Gels}} in {{Colloidal
  Systems}} with {{Short}}-{{Range Depletion Attraction}} and {{Screened
  Electrostatic Repulsion}}},}\ }\href {\doibase 10.1021/jp052683g} {\bibfield
  {journal} {\bibinfo  {journal} {The Journal of Physical Chemistry B}\
  }\textbf {\bibinfo {volume} {109}},\ \bibinfo {pages} {21942--21953}
  (\bibinfo {year} {2005})}\BibitemShut {NoStop}%
\bibitem [{\citenamefont {Malins}\ \emph {et~al.}(2011)\citenamefont {Malins},
  \citenamefont {Williams}, \citenamefont {Eggers}, \citenamefont {Tanaka},\
  and\ \citenamefont {Royall}}]{malins2011}%
  \BibitemOpen
  \bibfield  {author} {\bibinfo {author} {\bibfnamefont {A.}~\bibnamefont
  {Malins}}, \bibinfo {author} {\bibfnamefont {S.~R.}\ \bibnamefont
  {Williams}}, \bibinfo {author} {\bibfnamefont {J.}~\bibnamefont {Eggers}},
  \bibinfo {author} {\bibfnamefont {H.}~\bibnamefont {Tanaka}}, \ and\ \bibinfo
  {author} {\bibfnamefont {C.~P.}\ \bibnamefont {Royall}},\ }\bibfield  {title}
  {\enquote {\bibinfo {title} {The effect of inter-cluster interactions on the
  structure of colloidal clusters},}\ }\href {\doibase
  10.1016/j.jnoncrysol.2010.08.021} {\bibfield  {journal} {\bibinfo  {journal}
  {Journal of Non-Crystalline Solids}\ }\textbf {\bibinfo {volume} {357}},\
  \bibinfo {pages} {760--766} (\bibinfo {year} {2011})}\BibitemShut {NoStop}%
\bibitem [{\citenamefont {Klix}, \citenamefont {Royall},\ and\ \citenamefont
  {Tanaka}(2010)}]{klix2010}%
  \BibitemOpen
  \bibfield  {author} {\bibinfo {author} {\bibfnamefont {C.~L.}\ \bibnamefont
  {Klix}}, \bibinfo {author} {\bibfnamefont {C.~P.}\ \bibnamefont {Royall}}, \
  and\ \bibinfo {author} {\bibfnamefont {H.}~\bibnamefont {Tanaka}},\
  }\bibfield  {title} {\enquote {\bibinfo {title} {Structural and {{Dynamical
  Features}} of {{Multiple Metastable Glassy States}} in a {{Colloidal System}}
  with {{Competing Interactions}}},}\ }\href {\doibase
  10.1103/PhysRevLett.104.165702} {\bibfield  {journal} {\bibinfo  {journal}
  {Physical Review Letters}\ }\textbf {\bibinfo {volume} {104}},\ \bibinfo
  {pages} {165702} (\bibinfo {year} {2010})}\BibitemShut {NoStop}%
\bibitem [{\citenamefont {{van Gruijthuijsen}}\ \emph
  {et~al.}(2018)\citenamefont {{van Gruijthuijsen}}, \citenamefont
  {{Obiols-Rabasa}}, \citenamefont {Schurtenberger}, \citenamefont {Bouwman},\
  and\ \citenamefont {Stradner}}]{vangruijthuijsen2018}%
  \BibitemOpen
  \bibfield  {author} {\bibinfo {author} {\bibfnamefont {K.}~\bibnamefont {{van
  Gruijthuijsen}}}, \bibinfo {author} {\bibfnamefont {M.}~\bibnamefont
  {{Obiols-Rabasa}}}, \bibinfo {author} {\bibfnamefont {P.}~\bibnamefont
  {Schurtenberger}}, \bibinfo {author} {\bibfnamefont {W.~G.}\ \bibnamefont
  {Bouwman}}, \ and\ \bibinfo {author} {\bibfnamefont {A.}~\bibnamefont
  {Stradner}},\ }\bibfield  {title} {\enquote {\bibinfo {title} {The extended
  law of corresponding states when attractions meet repulsions},}\ }\href
  {\doibase 10.1039/C8SM00160J} {\bibfield  {journal} {\bibinfo  {journal}
  {Soft Matter}\ }\textbf {\bibinfo {volume} {14}},\ \bibinfo {pages}
  {3704--3715} (\bibinfo {year} {2018})}\BibitemShut {NoStop}%
\bibitem [{\citenamefont {Shukla}\ \emph {et~al.}(2008)\citenamefont {Shukla},
  \citenamefont {Mylonas}, \citenamefont {Di~Cola}, \citenamefont {Finet},
  \citenamefont {Timmins}, \citenamefont {Narayanan},\ and\ \citenamefont
  {Svergun}}]{shukla2008}%
  \BibitemOpen
  \bibfield  {author} {\bibinfo {author} {\bibfnamefont {A.}~\bibnamefont
  {Shukla}}, \bibinfo {author} {\bibfnamefont {E.}~\bibnamefont {Mylonas}},
  \bibinfo {author} {\bibfnamefont {E.}~\bibnamefont {Di~Cola}}, \bibinfo
  {author} {\bibfnamefont {S.}~\bibnamefont {Finet}}, \bibinfo {author}
  {\bibfnamefont {P.}~\bibnamefont {Timmins}}, \bibinfo {author} {\bibfnamefont
  {T.}~\bibnamefont {Narayanan}}, \ and\ \bibinfo {author} {\bibfnamefont
  {D.~I.}\ \bibnamefont {Svergun}},\ }\bibfield  {title} {\enquote {\bibinfo
  {title} {Absence of equilibrium cluster phase in concentrated lysozyme
  solutions},}\ }\href {\doibase 10.1073/pnas.0711928105} {\bibfield  {journal}
  {\bibinfo  {journal} {Proceedings of the National Academy of Sciences}\
  }\textbf {\bibinfo {volume} {105}},\ \bibinfo {pages} {5075--5080} (\bibinfo
  {year} {2008})}\BibitemShut {NoStop}%
\bibitem [{\citenamefont {Klix}\ \emph {et~al.}(2013)\citenamefont {Klix},
  \citenamefont {Murata}, \citenamefont {Tanaka}, \citenamefont {Williams},
  \citenamefont {Malins},\ and\ \citenamefont {Royall}}]{klix2013}%
  \BibitemOpen
  \bibfield  {author} {\bibinfo {author} {\bibfnamefont {C.~L.}\ \bibnamefont
  {Klix}}, \bibinfo {author} {\bibfnamefont {K.-i.}\ \bibnamefont {Murata}},
  \bibinfo {author} {\bibfnamefont {H.}~\bibnamefont {Tanaka}}, \bibinfo
  {author} {\bibfnamefont {S.~R.}\ \bibnamefont {Williams}}, \bibinfo {author}
  {\bibfnamefont {A.}~\bibnamefont {Malins}}, \ and\ \bibinfo {author}
  {\bibfnamefont {C.~P.}\ \bibnamefont {Royall}},\ }\bibfield  {title}
  {\enquote {\bibinfo {title} {Novel kinetic trapping in charged colloidal
  clusters due to self-induced surface charge organization},}\ }\href@noop {}
  {\bibfield  {journal} {\bibinfo  {journal} {Scientific reports}\ }\textbf
  {\bibinfo {volume} {3}},\ \bibinfo {pages} {1--6} (\bibinfo {year}
  {2013})}\BibitemShut {NoStop}%
\bibitem [{\citenamefont {Likos}(2001)}]{likos2001}%
  \BibitemOpen
  \bibfield  {author} {\bibinfo {author} {\bibfnamefont {C.~N.}\ \bibnamefont
  {Likos}},\ }\bibfield  {title} {\enquote {\bibinfo {title} {Effective
  interactions in soft condensed matter physics},}\ }\href@noop {} {\bibfield
  {journal} {\bibinfo  {journal} {Physics Reports}\ }\textbf {\bibinfo {volume}
  {348}},\ \bibinfo {pages} {267--439} (\bibinfo {year} {2001})}\BibitemShut
  {NoStop}%
\bibitem [{\citenamefont {Baaden}\ and\ \citenamefont
  {Marrink}(2013)}]{baaden2013}%
  \BibitemOpen
  \bibfield  {author} {\bibinfo {author} {\bibfnamefont {M.}~\bibnamefont
  {Baaden}}\ and\ \bibinfo {author} {\bibfnamefont {S.~J.}\ \bibnamefont
  {Marrink}},\ }\bibfield  {title} {\enquote {\bibinfo {title} {Coarse-grain
  modelling of protein\textendash protein interactions},}\ }\href {\doibase
  10.1016/j.sbi.2013.09.004} {\bibfield  {journal} {\bibinfo  {journal}
  {Current Opinion in Structural Biology}\ }\textbf {\bibinfo {volume} {23}},\
  \bibinfo {pages} {878--886} (\bibinfo {year} {2013})}\BibitemShut {NoStop}%
\bibitem [{\citenamefont {Liu}, \citenamefont {Kumar},\ and\ \citenamefont
  {Douglas}(2009)}]{liu2009}%
  \BibitemOpen
  \bibfield  {author} {\bibinfo {author} {\bibfnamefont {H.}~\bibnamefont
  {Liu}}, \bibinfo {author} {\bibfnamefont {S.~K.}\ \bibnamefont {Kumar}}, \
  and\ \bibinfo {author} {\bibfnamefont {J.~F.}\ \bibnamefont {Douglas}},\
  }\bibfield  {title} {\enquote {\bibinfo {title} {Self-assembly-induced
  protein crystallization},}\ }\href@noop {} {\bibfield  {journal} {\bibinfo
  {journal} {Physical review letters}\ }\textbf {\bibinfo {volume} {103}},\
  \bibinfo {pages} {018101} (\bibinfo {year} {2009})}\BibitemShut {NoStop}%
\bibitem [{\citenamefont {Fusco}\ \emph {et~al.}(2014)\citenamefont {Fusco},
  \citenamefont {Headd}, \citenamefont {De~Simone}, \citenamefont {Wang},\ and\
  \citenamefont {Charbonneau}}]{fusco2014}%
  \BibitemOpen
  \bibfield  {author} {\bibinfo {author} {\bibfnamefont {D.}~\bibnamefont
  {Fusco}}, \bibinfo {author} {\bibfnamefont {J.~J.}\ \bibnamefont {Headd}},
  \bibinfo {author} {\bibfnamefont {A.}~\bibnamefont {De~Simone}}, \bibinfo
  {author} {\bibfnamefont {J.}~\bibnamefont {Wang}}, \ and\ \bibinfo {author}
  {\bibfnamefont {P.}~\bibnamefont {Charbonneau}},\ }\bibfield  {title}
  {\enquote {\bibinfo {title} {Characterizing protein crystal contacts and
  their role in crystallization: rubredoxin as a case study},}\ }\href@noop {}
  {\bibfield  {journal} {\bibinfo  {journal} {Soft matter}\ }\textbf {\bibinfo
  {volume} {10}},\ \bibinfo {pages} {290--302} (\bibinfo {year}
  {2014})}\BibitemShut {NoStop}%
\bibitem [{\citenamefont {Doye}\ and\ \citenamefont {Poon}(2006)}]{doye2006}%
  \BibitemOpen
  \bibfield  {author} {\bibinfo {author} {\bibfnamefont {J.~P.}\ \bibnamefont
  {Doye}}\ and\ \bibinfo {author} {\bibfnamefont {W.~C.}\ \bibnamefont
  {Poon}},\ }\bibfield  {title} {\enquote {\bibinfo {title} {Protein
  crystallization in vivo},}\ }\href@noop {} {\bibfield  {journal} {\bibinfo
  {journal} {Current opinion in colloid \& interface science}\ }\textbf
  {\bibinfo {volume} {11}},\ \bibinfo {pages} {40--46} (\bibinfo {year}
  {2006})}\BibitemShut {NoStop}%
\bibitem [{\citenamefont {Altan}\ \emph {et~al.}(2018)\citenamefont {Altan},
  \citenamefont {Fusco}, \citenamefont {Afonine},\ and\ \citenamefont
  {Charbonneau}}]{altan2018}%
  \BibitemOpen
  \bibfield  {author} {\bibinfo {author} {\bibfnamefont {I.}~\bibnamefont
  {Altan}}, \bibinfo {author} {\bibfnamefont {D.}~\bibnamefont {Fusco}},
  \bibinfo {author} {\bibfnamefont {P.~V.}\ \bibnamefont {Afonine}}, \ and\
  \bibinfo {author} {\bibfnamefont {P.}~\bibnamefont {Charbonneau}},\
  }\bibfield  {title} {\enquote {\bibinfo {title} {Learning about biomolecular
  solvation from water in protein crystals},}\ }\href@noop {} {\bibfield
  {journal} {\bibinfo  {journal} {The Journal of Physical Chemistry B}\
  }\textbf {\bibinfo {volume} {122}},\ \bibinfo {pages} {2475--2486} (\bibinfo
  {year} {2018})}\BibitemShut {NoStop}%
\bibitem [{\citenamefont {James}, \citenamefont {Quinn},\ and\ \citenamefont
  {McManus}(2015)}]{james2015}%
  \BibitemOpen
  \bibfield  {author} {\bibinfo {author} {\bibfnamefont {S.}~\bibnamefont
  {James}}, \bibinfo {author} {\bibfnamefont {M.~K.}\ \bibnamefont {Quinn}}, \
  and\ \bibinfo {author} {\bibfnamefont {J.~J.}\ \bibnamefont {McManus}},\
  }\bibfield  {title} {\enquote {\bibinfo {title} {The self assembly of
  proteins; probing patchy protein interactions},}\ }\href@noop {} {\bibfield
  {journal} {\bibinfo  {journal} {Physical Chemistry Chemical Physics}\
  }\textbf {\bibinfo {volume} {17}},\ \bibinfo {pages} {5413--5420} (\bibinfo
  {year} {2015})}\BibitemShut {NoStop}%
\bibitem [{\citenamefont {Li}\ \emph {et~al.}(2012)\citenamefont {Li},
  \citenamefont {Shi}, \citenamefont {An},\ and\ \citenamefont
  {Huang}}]{li2012}%
  \BibitemOpen
  \bibfield  {author} {\bibinfo {author} {\bibfnamefont {Y.}~\bibnamefont
  {Li}}, \bibinfo {author} {\bibfnamefont {T.}~\bibnamefont {Shi}}, \bibinfo
  {author} {\bibfnamefont {L.}~\bibnamefont {An}}, \ and\ \bibinfo {author}
  {\bibfnamefont {Q.}~\bibnamefont {Huang}},\ }\bibfield  {title} {\enquote
  {\bibinfo {title} {Monte {{Carlo Simulation}} on {{Complex Formation}} of
  {{Proteins}} and {{Polysaccharides}}},}\ }\href {\doibase 10.1021/jp206527p}
  {\bibfield  {journal} {\bibinfo  {journal} {The Journal of Physical Chemistry
  B}\ }\textbf {\bibinfo {volume} {116}},\ \bibinfo {pages} {3045--3053}
  (\bibinfo {year} {2012})}\BibitemShut {NoStop}%
\bibitem [{\citenamefont {Gnan}, \citenamefont {Sciortino},\ and\ \citenamefont
  {Zaccarelli}(2019)}]{gnan2019}%
  \BibitemOpen
  \bibfield  {author} {\bibinfo {author} {\bibfnamefont {N.}~\bibnamefont
  {Gnan}}, \bibinfo {author} {\bibfnamefont {F.}~\bibnamefont {Sciortino}}, \
  and\ \bibinfo {author} {\bibfnamefont {E.}~\bibnamefont {Zaccarelli}},\
  }\bibfield  {title} {\enquote {\bibinfo {title} {Patchy particle models to
  understand protein phase behavior},}\ }in\ \href@noop {} {\emph {\bibinfo
  {booktitle} {Protein Self-Assembly}}}\ (\bibinfo  {publisher} {Springer},\
  \bibinfo {year} {2019})\ pp.\ \bibinfo {pages} {187--208}\BibitemShut
  {NoStop}%
\bibitem [{\citenamefont {Cai}\ and\ \citenamefont {Sweeney}(2018)}]{cai2018}%
  \BibitemOpen
  \bibfield  {author} {\bibinfo {author} {\bibfnamefont {J.}~\bibnamefont
  {Cai}}\ and\ \bibinfo {author} {\bibfnamefont {A.~M.}\ \bibnamefont
  {Sweeney}},\ }\bibfield  {title} {\enquote {\bibinfo {title} {The proof is in
  the pidan: generalizing proteins as patchy particles},}\ }\href@noop {}
  {\bibfield  {journal} {\bibinfo  {journal} {ACS central science}\ }\textbf
  {\bibinfo {volume} {4}},\ \bibinfo {pages} {840--853} (\bibinfo {year}
  {2018})}\BibitemShut {NoStop}%
\bibitem [{\citenamefont {Asakura}\ and\ \citenamefont
  {Oosawa}(1954)}]{asakura1954}%
  \BibitemOpen
  \bibfield  {author} {\bibinfo {author} {\bibfnamefont {S.}~\bibnamefont
  {Asakura}}\ and\ \bibinfo {author} {\bibfnamefont {F.}~\bibnamefont
  {Oosawa}},\ }\bibfield  {title} {\enquote {\bibinfo {title} {On interaction
  between two bodies immersed in a solution of macromolecules},}\ }\href
  {\doibase 10.1063/1.1740347} {\bibfield  {journal} {\bibinfo  {journal} {The
  Journal of Chemical Physics}\ }\textbf {\bibinfo {volume} {22}},\ \bibinfo
  {pages} {1255--1256} (\bibinfo {year} {1954})}\BibitemShut {NoStop}%
\bibitem [{\citenamefont {Poon}(2002)}]{poon2002}%
  \BibitemOpen
  \bibfield  {author} {\bibinfo {author} {\bibfnamefont {W.}~\bibnamefont
  {Poon}},\ }\bibfield  {title} {\enquote {\bibinfo {title} {The physics of a
  model colloid--polymer mixture},}\ }\href@noop {} {\bibfield  {journal}
  {\bibinfo  {journal} {Journal of Physics: Condensed Matter}\ }\textbf
  {\bibinfo {volume} {14}},\ \bibinfo {pages} {R859--R880} (\bibinfo {year}
  {2002})}\BibitemShut {NoStop}%
\bibitem [{\citenamefont {Zaccarelli}(2007)}]{zaccarelli2007}%
  \BibitemOpen
  \bibfield  {author} {\bibinfo {author} {\bibfnamefont {E.}~\bibnamefont
  {Zaccarelli}},\ }\bibfield  {title} {\enquote {\bibinfo {title} {Colloidal
  gels: Equilibrium and non-equilibrium routes},}\ }\href {\doibase
  10.1088/0953-8984/19/32/323101} {\bibfield  {journal} {\bibinfo  {journal}
  {Journal of Physics: Condensed Matter}\ }\textbf {\bibinfo {volume} {19}},\
  \bibinfo {pages} {323101} (\bibinfo {year} {2007})}\BibitemShut {NoStop}%
\bibitem [{\citenamefont {Royall}\ \emph {et~al.}(2021)\citenamefont {Royall},
  \citenamefont {Faers}, \citenamefont {Fussell},\ and\ \citenamefont
  {Hallett}}]{royall2021}%
  \BibitemOpen
  \bibfield  {author} {\bibinfo {author} {\bibfnamefont {C.~P.}\ \bibnamefont
  {Royall}}, \bibinfo {author} {\bibfnamefont {M.~A.}\ \bibnamefont {Faers}},
  \bibinfo {author} {\bibfnamefont {S.}~\bibnamefont {Fussell}}, \ and\
  \bibinfo {author} {\bibfnamefont {J.}~\bibnamefont {Hallett}},\ }\bibfield
  {title} {\enquote {\bibinfo {title} {Real space analysis of colloidal gels:
  Triumphs, challenges and future directions},}\ }\href@noop {} {\bibfield
  {journal} {\bibinfo  {journal} {submitted to J. Phys.: Condens. Matter}\ }
  (\bibinfo {year} {2021})}\BibitemShut {NoStop}%
\bibitem [{\citenamefont {Asherie}, \citenamefont {Lomakin},\ and\
  \citenamefont {Benedek}(1996)}]{asherie1996}%
  \BibitemOpen
  \bibfield  {author} {\bibinfo {author} {\bibfnamefont {N.}~\bibnamefont
  {Asherie}}, \bibinfo {author} {\bibfnamefont {A.}~\bibnamefont {Lomakin}}, \
  and\ \bibinfo {author} {\bibfnamefont {G.~B.}\ \bibnamefont {Benedek}},\
  }\bibfield  {title} {\enquote {\bibinfo {title} {Phase diagram of colloidal
  solutions},}\ }\href@noop {} {\bibfield  {journal} {\bibinfo  {journal}
  {Physical review letters}\ }\textbf {\bibinfo {volume} {77}},\ \bibinfo
  {pages} {4832} (\bibinfo {year} {1996})}\BibitemShut {NoStop}%
\bibitem [{\citenamefont {Vivares}\ \emph {et~al.}(2002)\citenamefont
  {Vivares}, \citenamefont {Belloni}, \citenamefont {Tardieu},\ and\
  \citenamefont {Bonnete}}]{vivares2002}%
  \BibitemOpen
  \bibfield  {author} {\bibinfo {author} {\bibfnamefont {D.}~\bibnamefont
  {Vivares}}, \bibinfo {author} {\bibfnamefont {L.}~\bibnamefont {Belloni}},
  \bibinfo {author} {\bibfnamefont {A.}~\bibnamefont {Tardieu}}, \ and\
  \bibinfo {author} {\bibfnamefont {F.}~\bibnamefont {Bonnete}},\ }\bibfield
  {title} {\enquote {\bibinfo {title} {Catching the peg-induced attractive
  interaction between proteins},}\ }\href@noop {} {\bibfield  {journal}
  {\bibinfo  {journal} {The European Physical Journal E}\ }\textbf {\bibinfo
  {volume} {9}},\ \bibinfo {pages} {15--25} (\bibinfo {year}
  {2002})}\BibitemShut {NoStop}%
\bibitem [{\citenamefont {Bolhuis}, \citenamefont {Meijer},\ and\ \citenamefont
  {Louis}(2003)}]{bolhuis2003}%
  \BibitemOpen
  \bibfield  {author} {\bibinfo {author} {\bibfnamefont {P.~G.}\ \bibnamefont
  {Bolhuis}}, \bibinfo {author} {\bibfnamefont {E.~J.}\ \bibnamefont {Meijer}},
  \ and\ \bibinfo {author} {\bibfnamefont {A.~A.}\ \bibnamefont {Louis}},\
  }\bibfield  {title} {\enquote {\bibinfo {title} {Colloid-{{Polymer Mixtures}}
  in the {{Protein Limit}}},}\ }\href {\doibase 10.1103/PhysRevLett.90.068304}
  {\bibfield  {journal} {\bibinfo  {journal} {Physical Review Letters}\
  }\textbf {\bibinfo {volume} {90}},\ \bibinfo {pages} {068304} (\bibinfo
  {year} {2003})}\BibitemShut {NoStop}%
\bibitem [{\citenamefont {Mutch}, \citenamefont {van Duijneveldt},\ and\
  \citenamefont {Eastoe}(2007)}]{mutch2007}%
  \BibitemOpen
  \bibfield  {author} {\bibinfo {author} {\bibfnamefont {K.~J.}\ \bibnamefont
  {Mutch}}, \bibinfo {author} {\bibfnamefont {J.~S.}\ \bibnamefont {van
  Duijneveldt}}, \ and\ \bibinfo {author} {\bibfnamefont {J.}~\bibnamefont
  {Eastoe}},\ }\bibfield  {title} {\enquote {\bibinfo {title} {Colloid--polymer
  mixtures in the protein limit},}\ }\href@noop {} {\bibfield  {journal}
  {\bibinfo  {journal} {Soft Matter}\ }\textbf {\bibinfo {volume} {3}},\
  \bibinfo {pages} {155--167} (\bibinfo {year} {2007})}\BibitemShut {NoStop}%
\bibitem [{\citenamefont {Mutch}\ \emph {et~al.}(2008)\citenamefont {Mutch},
  \citenamefont {van Duijneveldt}, \citenamefont {Eastoe}, \citenamefont
  {Grillo},\ and\ \citenamefont {Heenan}}]{mutch2008}%
  \BibitemOpen
  \bibfield  {author} {\bibinfo {author} {\bibfnamefont {K.~J.}\ \bibnamefont
  {Mutch}}, \bibinfo {author} {\bibfnamefont {J.~S.}\ \bibnamefont {van
  Duijneveldt}}, \bibinfo {author} {\bibfnamefont {J.}~\bibnamefont {Eastoe}},
  \bibinfo {author} {\bibfnamefont {I.}~\bibnamefont {Grillo}}, \ and\ \bibinfo
  {author} {\bibfnamefont {R.~K.}\ \bibnamefont {Heenan}},\ }\bibfield  {title}
  {\enquote {\bibinfo {title} {Small-angle neutron scattering study of
  microemulsion- polymer mixtures in the protein limit},}\ }\href@noop {}
  {\bibfield  {journal} {\bibinfo  {journal} {Langmuir}\ }\textbf {\bibinfo
  {volume} {24}},\ \bibinfo {pages} {3053--3060} (\bibinfo {year}
  {2008})}\BibitemShut {NoStop}%
\bibitem [{\citenamefont {Atha}\ and\ \citenamefont {Ingham}(1981)}]{atha1981}%
  \BibitemOpen
  \bibfield  {author} {\bibinfo {author} {\bibfnamefont {D.}~\bibnamefont
  {Atha}}\ and\ \bibinfo {author} {\bibfnamefont {K.}~\bibnamefont {Ingham}},\
  }\bibfield  {title} {\enquote {\bibinfo {title} {Mechanism of precipitation
  of proteins by polyethylene glycols. {{Analysis}} in terms of excluded
  volume.}}\ }\href {\doibase 10.1016/S0021-9258(18)43240-1} {\bibfield
  {journal} {\bibinfo  {journal} {Journal of Biological Chemistry}\ }\textbf
  {\bibinfo {volume} {256}},\ \bibinfo {pages} {12108--12117} (\bibinfo {year}
  {1981})}\BibitemShut {NoStop}%
\bibitem [{\citenamefont {Ha{\v{s}}ek}(2006)}]{hasek2006}%
  \BibitemOpen
  \bibfield  {author} {\bibinfo {author} {\bibfnamefont {J.}~\bibnamefont
  {Ha{\v{s}}ek}},\ }\bibfield  {title} {\enquote {\bibinfo {title} {Poly
  (ethylene glycol) interactions with proteins},}\ }\href@noop {} {\bibfield
  {journal} {\bibinfo  {journal} {Z. Kristallogr. Suppl}\ }\textbf {\bibinfo
  {volume} {23}},\ \bibinfo {pages} {613--618} (\bibinfo {year}
  {2006})}\BibitemShut {NoStop}%
\bibitem [{\citenamefont {Shkel}, \citenamefont {Knowles},\ and\ \citenamefont
  {Record}(2015)}]{Shkel2015}%
  \BibitemOpen
  \bibfield  {author} {\bibinfo {author} {\bibfnamefont {I.~A.}\ \bibnamefont
  {Shkel}}, \bibinfo {author} {\bibfnamefont {D.~B.}\ \bibnamefont {Knowles}},
  \ and\ \bibinfo {author} {\bibfnamefont {M.~T.}\ \bibnamefont {Record}},\
  }\bibfield  {title} {\enquote {\bibinfo {title} {Separating chemical and
  excluded volume interactions of polyethylene glycols with native proteins:
  {{Comparison}} with {{PEG}} effects on {{DNA}} helix formation: {{Separating
  Chemical}} and {{Excluded Volume Effects}} of {{Polyethylene Glycol}}},}\
  }\href {\doibase 10.1002/bip.22662} {\bibfield  {journal} {\bibinfo
  {journal} {Biopolymers}\ }\textbf {\bibinfo {volume} {103}},\ \bibinfo
  {pages} {517--527} (\bibinfo {year} {2015})}\BibitemShut {NoStop}%
\bibitem [{\citenamefont {Durbin}\ and\ \citenamefont
  {Feher}(1996)}]{durbin1996}%
  \BibitemOpen
  \bibfield  {author} {\bibinfo {author} {\bibfnamefont {S.}~\bibnamefont
  {Durbin}}\ and\ \bibinfo {author} {\bibfnamefont {G.}~\bibnamefont {Feher}},\
  }\bibfield  {title} {\enquote {\bibinfo {title} {Protein crystallization},}\
  }\href@noop {} {\bibfield  {journal} {\bibinfo  {journal} {Annual review of
  physical chemistry}\ }\textbf {\bibinfo {volume} {47}},\ \bibinfo {pages}
  {171--204} (\bibinfo {year} {1996})}\BibitemShut {NoStop}%
\bibitem [{\citenamefont {Bolhuis}\ \emph {et~al.}(1997)\citenamefont
  {Bolhuis}, \citenamefont {Stroobants}, \citenamefont {Frenkel},\ and\
  \citenamefont {Lekkerkerker}}]{bolhuis1997}%
  \BibitemOpen
  \bibfield  {author} {\bibinfo {author} {\bibfnamefont {P.~G.}\ \bibnamefont
  {Bolhuis}}, \bibinfo {author} {\bibfnamefont {A.}~\bibnamefont {Stroobants}},
  \bibinfo {author} {\bibfnamefont {D.}~\bibnamefont {Frenkel}}, \ and\
  \bibinfo {author} {\bibfnamefont {H.~N.~W.}\ \bibnamefont {Lekkerkerker}},\
  }\bibfield  {title} {\enquote {\bibinfo {title} {Numerical study of the phase
  behavior of rodlike colloids with attractive interactions},}\ }\href
  {\doibase 10.1063/1.474508} {\bibfield  {journal} {\bibinfo  {journal} {The
  Journal of Chemical Physics}\ }\textbf {\bibinfo {volume} {107}},\ \bibinfo
  {pages} {1551--1564} (\bibinfo {year} {1997})}\BibitemShut {NoStop}%
\bibitem [{\citenamefont {Savenko}\ and\ \citenamefont
  {Dijkstra}(2006)}]{savenko2006}%
  \BibitemOpen
  \bibfield  {author} {\bibinfo {author} {\bibfnamefont {S.~V.}\ \bibnamefont
  {Savenko}}\ and\ \bibinfo {author} {\bibfnamefont {M.}~\bibnamefont
  {Dijkstra}},\ }\bibfield  {title} {\enquote {\bibinfo {title} {Phase behavior
  of a suspension of colloidal hard rods and nonadsorbing polymer},}\ }\href
  {\doibase 10.1063/1.2202853} {\bibfield  {journal} {\bibinfo  {journal} {The
  Journal of Chemical Physics}\ }\textbf {\bibinfo {volume} {124}},\ \bibinfo
  {pages} {234902} (\bibinfo {year} {2006})}\BibitemShut {NoStop}%
\bibitem [{\citenamefont {Lo~Verso}\ \emph {et~al.}(2006)\citenamefont
  {Lo~Verso}, \citenamefont {Vink}, \citenamefont {Pini},\ and\ \citenamefont
  {Reatto}}]{loverso2006}%
  \BibitemOpen
  \bibfield  {author} {\bibinfo {author} {\bibfnamefont {F.}~\bibnamefont
  {Lo~Verso}}, \bibinfo {author} {\bibfnamefont {R.~L.~C.}\ \bibnamefont
  {Vink}}, \bibinfo {author} {\bibfnamefont {D.}~\bibnamefont {Pini}}, \ and\
  \bibinfo {author} {\bibfnamefont {L.}~\bibnamefont {Reatto}},\ }\bibfield
  {title} {\enquote {\bibinfo {title} {Critical behavior in colloid-polymer
  mixtures: {{Theory}} and simulation},}\ }\href {\doibase
  10.1103/PhysRevE.73.061407} {\bibfield  {journal} {\bibinfo  {journal}
  {Physical Review E}\ }\textbf {\bibinfo {volume} {73}},\ \bibinfo {pages}
  {061407} (\bibinfo {year} {2006})}\BibitemShut {NoStop}%
\bibitem [{\citenamefont {Lekkerkerker}\ and\ \citenamefont
  {Tuinier}(2011)}]{lekkerkerker2011}%
  \BibitemOpen
  \bibfield  {author} {\bibinfo {author} {\bibfnamefont {H.~N.}\ \bibnamefont
  {Lekkerkerker}}\ and\ \bibinfo {author} {\bibfnamefont {R.}~\bibnamefont
  {Tuinier}},\ }\href {\doibase 10.1007/978-94-007-1223-2} {\emph {\bibinfo
  {title} {Colloids and the {{Depletion Interaction}}}}},\ \bibinfo {series}
  {Lecture {{Notes}} in {{Physics}}}, Vol.\ \bibinfo {volume} {833}\ (\bibinfo
  {publisher} {{Springer Netherlands}},\ \bibinfo {address} {{Dordrecht}},\
  \bibinfo {year} {2011})\BibitemShut {NoStop}%
\bibitem [{\citenamefont {Roosen-Runge}\ \emph {et~al.}(2013)\citenamefont
  {Roosen-Runge}, \citenamefont {Heck}, \citenamefont {Zhang}, \citenamefont
  {Kohlbacher},\ and\ \citenamefont {Schreiber}}]{roosen2013}%
  \BibitemOpen
  \bibfield  {author} {\bibinfo {author} {\bibfnamefont {F.}~\bibnamefont
  {Roosen-Runge}}, \bibinfo {author} {\bibfnamefont {B.~S.}\ \bibnamefont
  {Heck}}, \bibinfo {author} {\bibfnamefont {F.}~\bibnamefont {Zhang}},
  \bibinfo {author} {\bibfnamefont {O.}~\bibnamefont {Kohlbacher}}, \ and\
  \bibinfo {author} {\bibfnamefont {F.}~\bibnamefont {Schreiber}},\ }\bibfield
  {title} {\enquote {\bibinfo {title} {Interplay of ph and binding of
  multivalent metal ions: charge inversion and reentrant condensation in
  protein solutions},}\ }\href@noop {} {\bibfield  {journal} {\bibinfo
  {journal} {The Journal of Physical Chemistry B}\ }\textbf {\bibinfo {volume}
  {117}},\ \bibinfo {pages} {5777--5787} (\bibinfo {year} {2013})}\BibitemShut
  {NoStop}%
\bibitem [{\citenamefont {Tang}\ \emph {et~al.}(2018)\citenamefont {Tang},
  \citenamefont {Cecchi}, \citenamefont {Fracasso}, \citenamefont {Accardi},
  \citenamefont {Coutable-Pennarun}, \citenamefont {Mansy}, \citenamefont
  {Perriman}, \citenamefont {Anderson},\ and\ \citenamefont {Mann}}]{tang2018}%
  \BibitemOpen
  \bibfield  {author} {\bibinfo {author} {\bibfnamefont {T.~D.}\ \bibnamefont
  {Tang}}, \bibinfo {author} {\bibfnamefont {D.}~\bibnamefont {Cecchi}},
  \bibinfo {author} {\bibfnamefont {G.}~\bibnamefont {Fracasso}}, \bibinfo
  {author} {\bibfnamefont {D.}~\bibnamefont {Accardi}}, \bibinfo {author}
  {\bibfnamefont {A.}~\bibnamefont {Coutable-Pennarun}}, \bibinfo {author}
  {\bibfnamefont {S.~S.}\ \bibnamefont {Mansy}}, \bibinfo {author}
  {\bibfnamefont {A.~W.}\ \bibnamefont {Perriman}}, \bibinfo {author}
  {\bibfnamefont {J.~R.}\ \bibnamefont {Anderson}}, \ and\ \bibinfo {author}
  {\bibfnamefont {S.}~\bibnamefont {Mann}},\ }\bibfield  {title} {\enquote
  {\bibinfo {title} {Gene-mediated chemical communication in synthetic
  protocell communities},}\ }\href@noop {} {\bibfield  {journal} {\bibinfo
  {journal} {ACS synthetic biology}\ }\textbf {\bibinfo {volume} {7}},\
  \bibinfo {pages} {339--346} (\bibinfo {year} {2018})}\BibitemShut {NoStop}%
\bibitem [{\citenamefont {Kaishima}\ \emph {et~al.}(2016)\citenamefont
  {Kaishima}, \citenamefont {Ishii}, \citenamefont {Matsuno}, \citenamefont
  {Fukuda},\ and\ \citenamefont {Kondo}}]{kaishima2016}%
  \BibitemOpen
  \bibfield  {author} {\bibinfo {author} {\bibfnamefont {M.}~\bibnamefont
  {Kaishima}}, \bibinfo {author} {\bibfnamefont {J.}~\bibnamefont {Ishii}},
  \bibinfo {author} {\bibfnamefont {T.}~\bibnamefont {Matsuno}}, \bibinfo
  {author} {\bibfnamefont {N.}~\bibnamefont {Fukuda}}, \ and\ \bibinfo {author}
  {\bibfnamefont {A.}~\bibnamefont {Kondo}},\ }\bibfield  {title} {\enquote
  {\bibinfo {title} {Expression of varied {{GFPs}} in {{Saccharomyces}}
  cerevisiae: Codon optimization yields stronger than expected expression and
  fluorescence intensity},}\ }\href@noop {} {\bibfield  {journal} {\bibinfo
  {journal} {Scientific Reports}\ }\textbf {\bibinfo {volume} {6}},\ \bibinfo
  {pages} {35932} (\bibinfo {year} {2016})}\BibitemShut {NoStop}%
\bibitem [{\citenamefont {George}\ and\ \citenamefont
  {Wilson}(1994{\natexlab{b}})}]{crys1994}%
  \BibitemOpen
  \bibfield  {author} {\bibinfo {author} {\bibfnamefont {A.}~\bibnamefont
  {George}}\ and\ \bibinfo {author} {\bibfnamefont {W.~W.}\ \bibnamefont
  {Wilson}},\ }\bibfield  {title} {\enquote {\bibinfo {title} {Predicting
  protein crystallization from a dilute solution property},}\ }\href@noop {}
  {\bibfield  {journal} {\bibinfo  {journal} {Acta Crystallographica Section D:
  Biological Crystallography}\ }\textbf {\bibinfo {volume} {50}},\ \bibinfo
  {pages} {361--365} (\bibinfo {year} {1994}{\natexlab{b}})}\BibitemShut
  {NoStop}%
\bibitem [{\citenamefont {Doucet}\ \emph {et~al.}(2017)\citenamefont {Doucet},
  \citenamefont {Cho}, \citenamefont {Alina}, \citenamefont {Bakker},
  \citenamefont {Bouwman}, \citenamefont {Butler}, \citenamefont {Campbell},
  \citenamefont {Gonzales}, \citenamefont {Heenan}, \citenamefont {Jackson}
  \emph {et~al.}}]{sasview}%
  \BibitemOpen
  \bibfield  {author} {\bibinfo {author} {\bibfnamefont {M.}~\bibnamefont
  {Doucet}}, \bibinfo {author} {\bibfnamefont {J.~H.}\ \bibnamefont {Cho}},
  \bibinfo {author} {\bibfnamefont {G.}~\bibnamefont {Alina}}, \bibinfo
  {author} {\bibfnamefont {J.}~\bibnamefont {Bakker}}, \bibinfo {author}
  {\bibfnamefont {W.}~\bibnamefont {Bouwman}}, \bibinfo {author} {\bibfnamefont
  {P.}~\bibnamefont {Butler}}, \bibinfo {author} {\bibfnamefont
  {K.}~\bibnamefont {Campbell}}, \bibinfo {author} {\bibfnamefont
  {M.}~\bibnamefont {Gonzales}}, \bibinfo {author} {\bibfnamefont
  {R.}~\bibnamefont {Heenan}}, \bibinfo {author} {\bibfnamefont
  {A.}~\bibnamefont {Jackson}},  \emph {et~al.},\ }\href@noop {} {\enquote
  {\bibinfo {title} {Sasview},}\ } (\bibinfo {year} {2017}),\ \bibinfo {note}
  {\url{https://www.sasview.org/}}\BibitemShut {NoStop}%
\bibitem [{\citenamefont {Zhang}\ \emph {et~al.}(2007)\citenamefont {Zhang},
  \citenamefont {Skoda}, \citenamefont {Jacobs}, \citenamefont {Martin},
  \citenamefont {Martin},\ and\ \citenamefont {Schreiber}}]{zhang2007}%
  \BibitemOpen
  \bibfield  {author} {\bibinfo {author} {\bibfnamefont {F.}~\bibnamefont
  {Zhang}}, \bibinfo {author} {\bibfnamefont {M.~W.~A.}\ \bibnamefont {Skoda}},
  \bibinfo {author} {\bibfnamefont {R.~M.~J.}\ \bibnamefont {Jacobs}}, \bibinfo
  {author} {\bibfnamefont {R.~A.}\ \bibnamefont {Martin}}, \bibinfo {author}
  {\bibfnamefont {C.~M.}\ \bibnamefont {Martin}}, \ and\ \bibinfo {author}
  {\bibfnamefont {F.}~\bibnamefont {Schreiber}},\ }\bibfield  {title} {\enquote
  {\bibinfo {title} {Protein {{Interactions Studied}} by {{SAXS}}: {{Effect}}
  of {{Ionic Strength}} and {{Protein Concentration}} for {{BSA}} in {{Aqueous
  Solutions}}},}\ }\href@noop {} {\bibfield  {journal} {\bibinfo  {journal}
  {The Journal of Physical Chemistry B}\ }\textbf {\bibinfo {volume} {111}},\
  \bibinfo {pages} {251} (\bibinfo {year} {2007})}\BibitemShut {NoStop}%
\bibitem [{\citenamefont {Wolf}\ \emph {et~al.}(2014)\citenamefont {Wolf},
  \citenamefont {Roosen-Runge}, \citenamefont {Zhang}, \citenamefont {Roth},
  \citenamefont {Skoda}, \citenamefont {Jacobs}, \citenamefont {Sztucki},\ and\
  \citenamefont {Schreiber}}]{wolf2014}%
  \BibitemOpen
  \bibfield  {author} {\bibinfo {author} {\bibfnamefont {M.}~\bibnamefont
  {Wolf}}, \bibinfo {author} {\bibfnamefont {F.}~\bibnamefont {Roosen-Runge}},
  \bibinfo {author} {\bibfnamefont {F.}~\bibnamefont {Zhang}}, \bibinfo
  {author} {\bibfnamefont {R.}~\bibnamefont {Roth}}, \bibinfo {author}
  {\bibfnamefont {M.~W.}\ \bibnamefont {Skoda}}, \bibinfo {author}
  {\bibfnamefont {R.~M.}\ \bibnamefont {Jacobs}}, \bibinfo {author}
  {\bibfnamefont {M.}~\bibnamefont {Sztucki}}, \ and\ \bibinfo {author}
  {\bibfnamefont {F.}~\bibnamefont {Schreiber}},\ }\bibfield  {title} {\enquote
  {\bibinfo {title} {Effective interactions in protein--salt solutions
  approaching liquid--liquid phase separation},}\ }\href@noop {} {\bibfield
  {journal} {\bibinfo  {journal} {Journal of Molecular Liquids}\ }\textbf
  {\bibinfo {volume} {200}},\ \bibinfo {pages} {20--27} (\bibinfo {year}
  {2014})}\BibitemShut {NoStop}%
\bibitem [{\citenamefont {Singh}\ \emph {et~al.}(2019)\citenamefont {Singh},
  \citenamefont {Roche}, \citenamefont {Van~der Walle}, \citenamefont {Uddin},
  \citenamefont {Du}, \citenamefont {Warwicker}, \citenamefont {Pluen},\ and\
  \citenamefont {Curtis}}]{singh2019}%
  \BibitemOpen
  \bibfield  {author} {\bibinfo {author} {\bibfnamefont {P.}~\bibnamefont
  {Singh}}, \bibinfo {author} {\bibfnamefont {A.}~\bibnamefont {Roche}},
  \bibinfo {author} {\bibfnamefont {C.~F.}\ \bibnamefont {Van~der Walle}},
  \bibinfo {author} {\bibfnamefont {S.}~\bibnamefont {Uddin}}, \bibinfo
  {author} {\bibfnamefont {J.}~\bibnamefont {Du}}, \bibinfo {author}
  {\bibfnamefont {J.}~\bibnamefont {Warwicker}}, \bibinfo {author}
  {\bibfnamefont {A.}~\bibnamefont {Pluen}}, \ and\ \bibinfo {author}
  {\bibfnamefont {R.}~\bibnamefont {Curtis}},\ }\bibfield  {title} {\enquote
  {\bibinfo {title} {Determination of protein--protein interactions in a
  mixture of two monoclonal antibodies},}\ }\href@noop {} {\bibfield  {journal}
  {\bibinfo  {journal} {Molecular pharmaceutics}\ }\textbf {\bibinfo {volume}
  {16}},\ \bibinfo {pages} {4775--4786} (\bibinfo {year} {2019})}\BibitemShut
  {NoStop}%
\bibitem [{\citenamefont {Arpino}, \citenamefont {Rizkallah},\ and\
  \citenamefont {Jones}(2012)}]{arpino2012}%
  \BibitemOpen
  \bibfield  {author} {\bibinfo {author} {\bibfnamefont {J.~A.}\ \bibnamefont
  {Arpino}}, \bibinfo {author} {\bibfnamefont {P.~J.}\ \bibnamefont
  {Rizkallah}}, \ and\ \bibinfo {author} {\bibfnamefont {D.~D.}\ \bibnamefont
  {Jones}},\ }\bibfield  {title} {\enquote {\bibinfo {title} {Crystal structure
  of enhanced green fluorescent protein to 1.35 {\aa} resolution reveals
  alternative conformations for glu222},}\ }\href@noop {} {\bibfield  {journal}
  {\bibinfo  {journal} {PloS one}\ }\textbf {\bibinfo {volume} {7}},\ \bibinfo
  {pages} {e47132} (\bibinfo {year} {2012})}\BibitemShut {NoStop}%
\bibitem [{\citenamefont {Myatt}\ \emph {et~al.}(2017)\citenamefont {Myatt},
  \citenamefont {Hatter}, \citenamefont {Rogers}, \citenamefont {Terry},\ and\
  \citenamefont {Clifton}}]{myatt2017}%
  \BibitemOpen
  \bibfield  {author} {\bibinfo {author} {\bibfnamefont {D.~P.}\ \bibnamefont
  {Myatt}}, \bibinfo {author} {\bibfnamefont {L.}~\bibnamefont {Hatter}},
  \bibinfo {author} {\bibfnamefont {S.~E.}\ \bibnamefont {Rogers}}, \bibinfo
  {author} {\bibfnamefont {A.~E.}\ \bibnamefont {Terry}}, \ and\ \bibinfo
  {author} {\bibfnamefont {L.~A.}\ \bibnamefont {Clifton}},\ }\bibfield
  {title} {\enquote {\bibinfo {title} {Monomeric green fluorescent protein as a
  protein standard for small angle scattering},}\ }\href@noop {} {\bibfield
  {journal} {\bibinfo  {journal} {Biomedical Spectroscopy and Imaging}\
  }\textbf {\bibinfo {volume} {6}},\ \bibinfo {pages} {123--134} (\bibinfo
  {year} {2017})}\BibitemShut {NoStop}%
\bibitem [{\citenamefont {Piazza}(2014)}]{piazza2014}%
  \BibitemOpen
  \bibfield  {author} {\bibinfo {author} {\bibfnamefont {R.}~\bibnamefont
  {Piazza}},\ }\bibfield  {title} {\enquote {\bibinfo {title} {Settled and
  unsettled issues in particle settling},}\ }\href {\doibase
  10.1088/0034-4885/77/5/056602} {\bibfield  {journal} {\bibinfo  {journal}
  {Reports on Progress in Physics}\ }\textbf {\bibinfo {volume} {77}},\
  \bibinfo {pages} {056602} (\bibinfo {year} {2014})}\BibitemShut {NoStop}%
\bibitem [{\citenamefont {Poon}, \citenamefont {Weeks},\ and\ \citenamefont
  {Royall}(2012)}]{poon2012}%
  \BibitemOpen
  \bibfield  {author} {\bibinfo {author} {\bibfnamefont {W.~C.~K.}\
  \bibnamefont {Poon}}, \bibinfo {author} {\bibfnamefont {E.~R.}\ \bibnamefont
  {Weeks}}, \ and\ \bibinfo {author} {\bibfnamefont {C.~P.}\ \bibnamefont
  {Royall}},\ }\bibfield  {title} {\enquote {\bibinfo {title} {On measuring
  colloidal volume fractions},}\ }\href {\doibase 10.1039/CISM06083J}
  {\bibfield  {journal} {\bibinfo  {journal} {Soft Matter}\ }\textbf {\bibinfo
  {volume} {8}},\ \bibinfo {pages} {21--30} (\bibinfo {year}
  {2012})}\BibitemShut {NoStop}%
\bibitem [{\citenamefont {{ten Wolde}}, \citenamefont {{Ruiz-Montero}},\ and\
  \citenamefont {Frenkel}(1996)}]{tenwolde1996}%
  \BibitemOpen
  \bibfield  {author} {\bibinfo {author} {\bibfnamefont {P.-R.}\ \bibnamefont
  {{ten Wolde}}}, \bibinfo {author} {\bibfnamefont {M.~J.}\ \bibnamefont
  {{Ruiz-Montero}}}, \ and\ \bibinfo {author} {\bibfnamefont {D.}~\bibnamefont
  {Frenkel}},\ }\bibfield  {title} {\enquote {\bibinfo {title} {Simulation of
  homogeneous crystal nucleation close to coexistence},}\ }\href {\doibase
  10.1039/fd9960400093} {\bibfield  {journal} {\bibinfo  {journal} {Faraday
  Discussions}\ }\textbf {\bibinfo {volume} {104}},\ \bibinfo {pages} {93}
  (\bibinfo {year} {1996})}\BibitemShut {NoStop}%
\bibitem [{\citenamefont {Lekkerkerker}\ \emph {et~al.}(1992)\citenamefont
  {Lekkerkerker}, \citenamefont {Poon}, \citenamefont {Pusey}, \citenamefont
  {Stroobants},\ and\ \citenamefont {Warren}}]{lekkerkerker1992}%
  \BibitemOpen
  \bibfield  {author} {\bibinfo {author} {\bibfnamefont {H.~N.~W.}\
  \bibnamefont {Lekkerkerker}, \bibfnamefont {H.N.W.}}, \bibinfo {author}
  {\bibfnamefont {W.~C.~K.}\ \bibnamefont {Poon}}, \bibinfo {author}
  {\bibfnamefont {P.~N.}\ \bibnamefont {Pusey}}, \bibinfo {author}
  {\bibfnamefont {A.}~\bibnamefont {Stroobants}}, \ and\ \bibinfo {author}
  {\bibfnamefont {P.~B.}\ \bibnamefont {Warren}},\ }\bibfield  {title}
  {\enquote {\bibinfo {title} {Phase-behavior of colloid plus polymer
  mixtures},}\ }\href@noop {} {\bibfield  {journal} {\bibinfo  {journal}
  {Europhys. Lett.}\ }\textbf {\bibinfo {volume} {20}},\ \bibinfo {pages}
  {559--564} (\bibinfo {year} {1992})}\BibitemShut {NoStop}%
\bibitem [{\citenamefont {Dijkstra}, \citenamefont {Brader},\ and\
  \citenamefont {Evans}(1999)}]{dijkstra1999}%
  \BibitemOpen
  \bibfield  {author} {\bibinfo {author} {\bibfnamefont {M.}~\bibnamefont
  {Dijkstra}}, \bibinfo {author} {\bibfnamefont {J.~M.}\ \bibnamefont
  {Brader}}, \ and\ \bibinfo {author} {\bibfnamefont {R.}~\bibnamefont
  {Evans}},\ }\bibfield  {title} {\enquote {\bibinfo {title} {Phase behaviour
  and structure of model colloid-polymer mixtures},}\ }\href {\doibase
  10.1088/0953-8984/11/50/304} {\bibfield  {journal} {\bibinfo  {journal}
  {Journal of Physics: Condensed Matter}\ }\textbf {\bibinfo {volume} {11}},\
  \bibinfo {pages} {10079--10106} (\bibinfo {year} {1999})}\BibitemShut
  {NoStop}%
\bibitem [{\citenamefont {Fleer}\ and\ \citenamefont
  {Tuinier}(2008)}]{fleer2008}%
  \BibitemOpen
  \bibfield  {author} {\bibinfo {author} {\bibfnamefont {G.~J.}\ \bibnamefont
  {Fleer}}\ and\ \bibinfo {author} {\bibfnamefont {R.}~\bibnamefont
  {Tuinier}},\ }\bibfield  {title} {\enquote {\bibinfo {title} {Analytical
  phase diagrams for colloids and non-adsorbing polymer},}\ }\href {\doibase
  10.1016/j.cis.2008.07.001} {\bibfield  {journal} {\bibinfo  {journal}
  {Advances in Colloid and Interface Science}\ }\textbf {\bibinfo {volume}
  {143}},\ \bibinfo {pages} {1--47} (\bibinfo {year} {2008})}\BibitemShut
  {NoStop}%
\bibitem [{\citenamefont {Royall}, \citenamefont {Louis},\ and\ \citenamefont
  {Tanaka}(2007)}]{royall2007}%
  \BibitemOpen
  \bibfield  {author} {\bibinfo {author} {\bibfnamefont {C.~P.}\ \bibnamefont
  {Royall}}, \bibinfo {author} {\bibfnamefont {A.~A.}\ \bibnamefont {Louis}}, \
  and\ \bibinfo {author} {\bibfnamefont {H.}~\bibnamefont {Tanaka}},\
  }\bibfield  {title} {\enquote {\bibinfo {title} {Measuring colloidal
  interactions with confocal microscopy},}\ }\href@noop {} {\bibfield
  {journal} {\bibinfo  {journal} {The Journal of chemical physics}\ }\textbf
  {\bibinfo {volume} {127}},\ \bibinfo {pages} {044507} (\bibinfo {year}
  {2007})}\BibitemShut {NoStop}%
\bibitem [{Pep(2015)}]{PepCalc}%
  \BibitemOpen
  \href@noop {} {\enquote {\bibinfo {title} {Pepcalc.com--innovagen peptide
  property calculator},}\ } (\bibinfo {year} {Accessed: 23 Sept 2015}),\
  \bibinfo {note} {\url{http://pepcalc.com/}}\BibitemShut {NoStop}%
\bibitem [{\citenamefont {Royall}\ and\ \citenamefont
  {Malins}(2012)}]{royall2012}%
  \BibitemOpen
  \bibfield  {author} {\bibinfo {author} {\bibfnamefont {C.~P.}\ \bibnamefont
  {Royall}}\ and\ \bibinfo {author} {\bibfnamefont {A.}~\bibnamefont
  {Malins}},\ }\bibfield  {title} {\enquote {\bibinfo {title} {The role of
  quench rate in colloidal gels},}\ }\href {\doibase DOI: 10.1039/c2fd20041d}
  {\bibfield  {journal} {\bibinfo  {journal} {Faraday Discuss.}\ }\textbf
  {\bibinfo {volume} {158}},\ \bibinfo {pages} {301–311} (\bibinfo {year}
  {2012})}\BibitemShut {NoStop}%
\bibitem [{\citenamefont {Klotsa}\ and\ \citenamefont
  {Jack}(2011)}]{klotsa2011}%
  \BibitemOpen
  \bibfield  {author} {\bibinfo {author} {\bibfnamefont {D.}~\bibnamefont
  {Klotsa}}\ and\ \bibinfo {author} {\bibfnamefont {R.~L.}\ \bibnamefont
  {Jack}},\ }\bibfield  {title} {\enquote {\bibinfo {title} {Predicting the
  self-assembly of a model colloidal crystal},}\ }\href@noop {} {\bibfield
  {journal} {\bibinfo  {journal} {Soft Matter}\ }\textbf {\bibinfo {volume}
  {7}},\ \bibinfo {pages} {6294--6303} (\bibinfo {year} {2011})}\BibitemShut
  {NoStop}%
\bibitem [{\citenamefont {Kirkwood}\ \emph {et~al.}(2015)\citenamefont
  {Kirkwood}, \citenamefont {Hargreaves}, \citenamefont {O'Keefe},\ and\
  \citenamefont {Wilson}}]{kirkwood2015}%
  \BibitemOpen
  \bibfield  {author} {\bibinfo {author} {\bibfnamefont {J.}~\bibnamefont
  {Kirkwood}}, \bibinfo {author} {\bibfnamefont {D.}~\bibnamefont
  {Hargreaves}}, \bibinfo {author} {\bibfnamefont {S.}~\bibnamefont {O'Keefe}},
  \ and\ \bibinfo {author} {\bibfnamefont {J.}~\bibnamefont {Wilson}},\
  }\bibfield  {title} {\enquote {\bibinfo {title} {Using isoelectric point to
  determine the {{pH}} for initial protein crystallization trials},}\ }\href
  {\doibase 10.1093/bioinformatics/btv011} {\bibfield  {journal} {\bibinfo
  {journal} {Bioinformatics}\ }\textbf {\bibinfo {volume} {31}},\ \bibinfo
  {pages} {1444--1451} (\bibinfo {year} {2015})}\BibitemShut {NoStop}%
\bibitem [{\citenamefont {Prevost}\ \emph {et~al.}(1991)\citenamefont
  {Prevost}, \citenamefont {Wodak}, \citenamefont {Tidor},\ and\ \citenamefont
  {Karplus}}]{prevost1991}%
  \BibitemOpen
  \bibfield  {author} {\bibinfo {author} {\bibfnamefont {M.}~\bibnamefont
  {Prevost}}, \bibinfo {author} {\bibfnamefont {S.~J.}\ \bibnamefont {Wodak}},
  \bibinfo {author} {\bibfnamefont {B.}~\bibnamefont {Tidor}}, \ and\ \bibinfo
  {author} {\bibfnamefont {M.}~\bibnamefont {Karplus}},\ }\bibfield  {title}
  {\enquote {\bibinfo {title} {Contribution of the hydrophobic effect to
  protein stability: Analysis based on simulations of the {{Ile}}-96----{{Ala}}
  mutation in barnase.}}\ }\href {\doibase 10.1073/pnas.88.23.10880} {\bibfield
   {journal} {\bibinfo  {journal} {Proceedings of the National Academy of
  Sciences}\ }\textbf {\bibinfo {volume} {88}},\ \bibinfo {pages}
  {10880--10884} (\bibinfo {year} {1991})}\BibitemShut {NoStop}%
\bibitem [{\citenamefont {Quinn}, \citenamefont {James},\ and\ \citenamefont
  {McManus}(2019)}]{quinn2019}%
  \BibitemOpen
  \bibfield  {author} {\bibinfo {author} {\bibfnamefont {M.~K.}\ \bibnamefont
  {Quinn}}, \bibinfo {author} {\bibfnamefont {S.}~\bibnamefont {James}}, \ and\
  \bibinfo {author} {\bibfnamefont {J.~J.}\ \bibnamefont {McManus}},\
  }\bibfield  {title} {\enquote {\bibinfo {title} {Chemical modification alters
  protein--protein interactions and can lead to lower protein solubility},}\
  }\href@noop {} {\bibfield  {journal} {\bibinfo  {journal} {The Journal of
  Physical Chemistry B}\ }\textbf {\bibinfo {volume} {123}},\ \bibinfo {pages}
  {4373--4379} (\bibinfo {year} {2019})}\BibitemShut {NoStop}%
\bibitem [{\citenamefont {Louis}\ \emph {et~al.}(2002)\citenamefont {Louis},
  \citenamefont {Bolhuis}, \citenamefont {Meijer},\ and\ \citenamefont
  {Hansen}}]{louis2002}%
  \BibitemOpen
  \bibfield  {author} {\bibinfo {author} {\bibfnamefont {A.}~\bibnamefont
  {Louis}}, \bibinfo {author} {\bibfnamefont {P.}~\bibnamefont {Bolhuis}},
  \bibinfo {author} {\bibfnamefont {E.}~\bibnamefont {Meijer}}, \ and\ \bibinfo
  {author} {\bibfnamefont {J.}~\bibnamefont {Hansen}},\ }\bibfield  {title}
  {\enquote {\bibinfo {title} {Polymer induced depletion potentials in
  polymer-colloid mixtures},}\ }\href@noop {} {\bibfield  {journal} {\bibinfo
  {journal} {The Journal of chemical physics}\ }\textbf {\bibinfo {volume}
  {117}},\ \bibinfo {pages} {1893--1907} (\bibinfo {year} {2002})}\BibitemShut
  {NoStop}%
\bibitem [{\citenamefont {Curtis}\ and\ \citenamefont
  {Lue}(2006)}]{Curtis2006}%
  \BibitemOpen
  \bibfield  {author} {\bibinfo {author} {\bibfnamefont {R.}~\bibnamefont
  {Curtis}}\ and\ \bibinfo {author} {\bibfnamefont {L.}~\bibnamefont {Lue}},\
  }\bibfield  {title} {\enquote {\bibinfo {title} {A molecular approach to
  bioseparations: {{Protein}}\textendash protein and protein\textendash salt
  interactions},}\ }\href {\doibase 10.1016/j.ces.2005.04.007} {\bibfield
  {journal} {\bibinfo  {journal} {Chemical Engineering Science}\ }\textbf
  {\bibinfo {volume} {61}},\ \bibinfo {pages} {907--923} (\bibinfo {year}
  {2006})}\BibitemShut {NoStop}%
\end{thebibliography}
%

\end{document}